\documentstyle[12pt,epsfig]{article}

\catcode`\@=11
\def\@citex[#1]#2{%
\if@filesw \immediate \write \@auxout {\string \citation {#2}}\fi
\@tempcntb\m@ne \let\@h@ld\relax \def\@citea{}%
\@cite{%
  \@for \@citeb:=#2\do {%
    \@ifundefined {b@\@citeb}%
      {\@h@ld\@citea\@tempcntb\m@ne{\bf ?}%
      \@warning {Citation `\@citeb ' on page \thepage \space undefined}}%
      {\@tempcnta\@tempcntb \advance\@tempcnta\@ne%
      \@tempcntb\number\csname b@\@citeb \endcsname \relax%
      \ifnum\@tempcnta=\@tempcntb 
        \ifx\@h@ld\relax%
          \edef \@h@ld{\@citea\csname b@\@citeb\endcsname}%
        \else%
          \edef\@h@ld{\ifmmode{-}\else--\fi\csname b@\@citeb\endcsname}%
        \fi%
      \else
        \@h@ld\@citea\csname b@\@citeb \endcsname%
        \let\@h@ld\relax%
      \fi}%
    \def\@citea{,\penalty\@highpenalty\,}%
  }\@h@ld
}{#1}}

\def\@citeb#1#2{{[#1]\if@tempswa , #2\fi}}
\def\@citeu#1#2{{$^{#1}$\if@tempswa , #2\fi }}
\def\@citep#1#2{{#1\if@tempswa , #2\fi}}

\def\bcites{         
        \catcode`\@=11
        \let\@cite=\@citeb
        \catcode`\@=12
}

\def\upcites{         
        \catcode`\@=11
        \let\@cite=\@citeu
        \catcode`\@=12
}

\def\plaincites{      
        \catcode`\@=11
        \let\@cite=\@citep
        \catcode`\@=12
}

\newcount\hour
\newcount\minute
\newtoks\amorpm
\hour=\time\divide\hour by 60
\minute=\time{\multiply\hour by 60 \global\advance\minute by-\hour}
\edef\standardtime{{\ifnum\hour<12 \global\amorpm={am}%
        \else\global\amorpm={pm}\advance\hour by-12 \fi
        \ifnum\hour=0 \hour=12 \fi
        \number\hour:\ifnum\minute<10 0\fi\number\minute\the\amorpm}}
\edef\militarytime{\number\hour:\ifnum\minute<10 0\fi\number\minute}

\def\draftlabel#1{{\@bsphack\if@filesw {\let\thepage\relax
   \xdef\@gtempa{\write\@auxout{\string
      \newlabel{#1}{{\@currentlabel}{\thepage}}}}}\@gtempa
   \if@nobreak \ifvmode\nobreak\fi\fi\fi\@esphack}
        \gdef\@eqnlabel{#1}}
\def\@eqnlabel{}
\def\@vacuum{}
\def\marginnote#1{}
\def\draftmarginnote#1{\marginpar{\raggedright\scriptsize\tt#1}}
\def\draft{
        \pagestyle{plain}
        \overfullrule=2pt
        \oddsidemargin -.5truein
        \def\@oddhead{\sl \phantom{\today\quad\militarytime} \hfil
        \smash{\Large\sl DRAFT} \hfil \today\quad\militarytime}
        \let\@evenhead\@oddhead
        \let\label=\draftlabel
        \let\marginnote=\draftmarginnote
        \def\ps@empty{\let\@mkboth\@gobbletwo
        \def\@oddfoot{\hfil \smash{\Large\sl DRAFT} \hfil}
        \let\@evenfoot\@oddhead}
        \def\@eqnnum{(\theequation)\rlap{\kern\marginparsep\tt\@eqnlabel}%
        \global\let\@eqnlabel\@vacuum}  }

\def\blackfonts{
        \font\blackboard=msbm10 scaled\magstep1
        \font\blackboards=msbm8
        \font\blackboardss=msbm6
}

\def\nblack{            
        \def\ZZ{{Z \n{10} Z}}
        \def\NN{{N \n{14} N}}
        \def\CC{{C \n{11} C}}
        \def\RR{{R \n{11} R}}
        \def\QQ{{Q \n{12} Q}}
        \def\PP{{P \n{11} P}}
}

\def\prep{         
        \catcode`\@=11
        \input art10.sty
        \catcode`\@=12
        
        \let\small\null
        \def\blackfonts{
                \font\blackboard=msbm10
                \font\blackboards=msbm7
                \font\blackboardss=msbm5
        }
        \let\sl\it
        \twocolumn
        \sloppy
        \voffset=-2.54truecm
        \hoffset=-2.54truecm
        \flushbottom
        \parindent 1em
        \leftmargini 2em
        \leftmarginv .5em
        \leftmarginvi .5em
        \marginparwidth 48pt
        \marginparsep 10pt
        \setlength{\columnsep}{2truecm}
        \setlength{\textwidth}{25.4truecm}
        \setlength{\textheight}{17truecm}
        \baselineskip=16pt
        \oddsidemargin .18truein
        \evensidemargin .17truein
}

\def\eqalign#1{\null\,\vcenter{\openup\jot\m@th
  \ialign{\strut\hfil$\displaystyle{##}$&$\displaystyle{{}##}$\hfil
      \crcr#1\crcr}}\,}
\def\eqalignno#1{\displ@y \tabskip\centering
  \halign to\displaywidth{\hfil$\@lign\displaystyle{##}$\tabskip\z@skip
    &$\@lign\displaystyle{{}##}$\hfil\tabskip\centering
    &\llap{$\@lign##$}\tabskip\z@skip\crcr
    #1\crcr}}

\def\section{\@startsection {section}{1}{\z@}{3.ex plus 1ex minus
 .2ex}{2.ex plus .2ex}{\large\bf}}
\def\subsection{\@startsection{subsection}{2}{\z@}{2.75ex plus 1ex minus
 .2ex}{1.5ex plus .2ex}{\bf}}        

\def\appendix{{\newpage\section*{Appendix}}\let\appendix\section%
        {\setcounter{section}{0}
        \gdef\thesection{\Alph{section}}}\section}

\def\abstract{\if@twocolumn
\section*{Abstract}
\else 
\begin{center}
{\bf Abstract\vspace{-.5em}\vspace{0pt}}
\end{center}
\quotation
\fi}

\catcode`\@=12

\def\apriori{{\it a priori\/}}

\def\Kahler{K\"ahler}

\newcommand{\beq}{\begin{equation}}
\newcommand{\eeq}{\end{equation}}
\newcommand{\beqa}{\begin{eqnarray}}
\newcommand{\eeqa}{\end{eqnarray}}

\newcommand{\R}{{\bf R}}
\newcommand{\C}{{\bf C}}
\newcommand{\e}{{\rm e}}

\newcommand{\tilQ}{\widetilde{Q}}

\newcommand{\dd}{{\rm d}}
\newcommand{\elst}{{\ell_{\it st}}}
\newcommand{\elel}{{\ell_{11}}}
\newcommand{\gst}{{g_{\rm st}}}
\newcommand{\MT}{{$M$ theory}~}

\def\noj#1,#2,{{\bf #1} (19#2)\ }
\def\jou#1,#2,#3,{{\sl #1\/ }{\bf #2} (19#3)\ }
\def\ann#1,#2,{{\sl Ann.\ Physics\/ }{\bf #1} (19#2)\ }
\def\cmp#1,#2,{{\sl Comm.\ Math.\ Phys.\/ }{\bf #1} (19#2)\ }
\def\ma#1,#2,{{\sl Math.\ Ann.\/ }{\bf #1} (19#2)\ }
\def\ng#1,#2,{{\sl Nagoya.\ Math.\ J.\/ }{\bf #1} (19#2)\ }
\def\jd#1,#2,{{\sl J.\ Diff.\ Geom.\/ }{\bf #1} (19#2)\ }
\def\invm#1,#2,{{\sl Invent.\ Math.\/ }{\bf #1} (19#2)\ }
\def\cq#1,#2,{{\sl Class.\ Quantum Grav.\/ }{\bf #1} (19#2)\ }
\def\cqg#1,#2,{{\sl Class.\ Quantum Grav.\/ }{\bf #1} (19#2)\ }
\def\ijmp#1,#2,{{\sl Int.\ J.\ Mod.\ Phys.\/ }{\bf A#1} (19#2)\ }
\def\jmphy#1,#2,{{\sl J.\ Geom.\ Phys.\/ }{\bf #1} (19#2)\ }
\def\jams#1,#2,{{\sl J.\ Amer.\ Math.\ Soc.\/ }{\bf #1} (19#2)\ }
\def\grg#1,#2,{{\sl Gen.\ Rel.\ Grav.\/ }{\bf #1} (19#2)\ }
\def\mpl#1,#2,{{\sl Mod.\ Phys.\ Lett.\/ }{\bf A#1} (19#2)\ }
\def\nc#1,#2,{{\sl Nuovo Cim.\/ }{\bf #1} (19#2)\ }
\def\np#1,#2,{{\sl Nucl.\ Phys.\/ }{\bf B#1} (19#2)\ }
\def\pl#1,#2,{{\sl Phys.\ Lett.\/ }{\bf #1B} (19#2)\ }
\def\pla#1,#2,{{\sl Phys.\ Lett.\/ }{\bf #1A} (19#2)\ }
\def\pr#1,#2,{{\sl Phys.\ Rev.\/ }{\bf #1} (19#2)\ }
\def\prd#1,#2,{{\sl Phys.\ Rev.\/ }{\bf D#1} (19#2)\ }
\def\prl#1,#2,{{\sl Phys.\ Rev.\ Lett.\/ }{\bf #1} (19#2)\ }
\def\prp#1,#2,{{\sl Phys.\ Rept.\/ }{\bf #1C} (19#2)\ }
\def\ptp#1,#2,{{\sl Prog.\ Theor.\ Phys.\/ }{\bf #1} (19#2)\ }
\def\ptpsup#1,#2,{{\sl Prog.\ Theor.\ Phys.\/ Suppl.\/ }{\bf #1} (19#2)\ }
\def\rmp#1,#2,{{\sl Rev.\ Mod.\ Phys.\/ }{\bf #1} (19#2)\ }
\def\yadfiz#1,#2,#3[#4,#5]{{\sl Yad.\ Fiz.\/ }{\bf #1} (19#2) #3%
\ [{\sl Sov.\ J.\ Nucl.\ Phys.\/ }{\bf #4} (19#2) #5]}
\def\zh#1,#2,#3[#4,#5]{{\sl Zh.\ Exp.\ Theor.\ Fiz.\/ }{\bf #1} (19#2) #3%
\ [{\sl Sov.\ Phys.\ JETP\/ }{\bf #4} (19#2) #5]}

\def\beq{\begin{equation}}
\def\eeq{\end{equation}}
\def\beqar{\begin{eqnarray}}
\def\eeqar{\end{eqnarray}}
\def\non{\nonumber}

\newcommand{\be}{\begin{equation}}
\newcommand{\ee}{\end{equation}}
\newcommand{\bea}{\begin{eqnarray}}
\newcommand{\eea}{\end{eqnarray}}

\def\nfrac#1#2{{\displaystyle{\vphantom1\smash{\lower.5ex\hbox{\small$#1$}}%
        \over\vphantom1\smash{\raise.25ex\hbox{\small$#2$}}}}}

\def\p#1{\mskip#1mu}
\def\n#1{\mskip-#1mu}
\def\stop{\p6.}
\def\comma{\p6,}

\def\to{\rightarrow}

\def\lae{\mathrel{\mathop{\smash{\lower .5 ex \hbox{$\stackrel<\sim$}}}}}
\def\lae{\mathrel{\mathop{\smash{\lower .5 ex \hbox{$\stackrel>\sim$}}}}}

\def\pa{\partial}

\def\Tr{{\rm Tr}}
\def\l:{\mathopen{:}\,}
\def\r:{\,\mathclose{:}}

\def\vm{\vec{m}}

\catcode`\@=11
\def\theequation{\arabic{equation}}
\catcode`\@=12

\nblack
\bcites

\nblack

\catcode`\@=11
\def\theequation{\thesection.\arabic{equation}}
\@addtoreset{equation}{section}
\@addtoreset{footnote}{section}
\@addtoreset{footnote}{subsection}
\catcode`\@=12

\newcommand{\beqn}{\begin{equation}}
\newcommand{\eeqn}{\end{equation}}
\newcommand{\beqnarray}{\begin{eqnarray}}
\newcommand{\eeqnarray}{\end{eqnarray}}
\newcommand {\bear} [1] {\begin {array} {#1}}
\newcommand {\ear} {\end {array}}

\newcommand{\CP}{{\bf C}{\rm P}}

\newcommand {\beqarn} {\begin{eqnarray*}}
\newcommand {\eeqarn} {\end{eqnarray*}}

\newcommand {\vw} {\vec{w}}

\begin{document}

\begin{titlepage}

\begin{center}
\today
\hfill                  hep-th/9711143

\vskip 1.5 cm
{\large \bf K\"ahler Potential and Higher Derivative Terms
from M Theory Fivebrane}
\vskip 1 cm 
{Jan de Boer, Kentaro Hori, Hirosi Ooguri and Yaron Oz}\\
\vskip 0.5cm
{\sl Department of Physics,
University of California at Berkeley\\
366 Le\thinspace Conte Hall, Berkeley, CA 94720-7300, U.S.A.\\
and\\
Theoretical Physics Group, Mail Stop 50A--5101\\
Ernest Orlando Lawrence Berkeley National Laboratory\\
Berkeley, CA 94720, U.S.A.\\}

\end{center}

\vskip 0.5 cm
\begin{abstract}

The construction of four dimensional supersymmetric gauge theories 
via the fivebrane of M theory wrapped around a Riemann surface  has been
successfully applied to the computation of holomorphic quantities
of field 
theory.
In this paper we compute non-holomorphic quantities in the eleven
dimensional supergravity limit of M theory. 
While the {\Kahler} potential
on the Coulomb of $N=2$ theories
is correctly reproduced, 
higher derivative terms in the $N=2$ effective action
differ from what is expected for the four dimensional gauge theory.
 For the {\Kahler} potential of $N=1$ theories at
an abelian Coulomb phase,
the result again differs from
what is expected for the four-dimensional gauge theory.
Using a  gravitational back reaction method for the fivebrane
we compute the metric on the Higgs branch of $N=2$ gauge theories.
Here we find an agreement with the results
expected for the gauge theories.
A similar computation of the metric on $N=1$ Higgs branches 
yields information on the complex structure associated with the flavor
rotation in one case
and the classical metric in another.
We discuss what
four-dimensional field theory quantities
can be computed via the fivebrane in the supergravity limit
of M theory.

\end{abstract}

\end{titlepage}

\section{Introduction}

Many gauge field theory results in various dimensions
have been obtained in the last year by realizing them on the
worldvolume of branes.
Another method applied to the study of gauge theories is
geometric engineering \cite{KV,KKLMV,KLMVW}.
In this paper we will be interested in studying four dimensional
gauge theories using the first method.

Webs of intersecting branes as a tool for studying gauge theories
with reduced number of supersymmetries have been introduced in
\cite{hw}. Such a web of intersecting branes
of Type IIA string theory describing $N=2$
gauge theories in four dimensions can be realized by a single
fivebrane of M theory wrapping a Riemann surface
\cite{KLMVW,witten1}. 
The Riemann surface is
the Seiberg-Witten curve \cite{sw1} and therefore the  fivebrane
configuration
encodes the structure of the moduli space of vacua.
Similar  webs of intersecting branes
of Type IIA string theory describing $N=1$
gauge theories in four dimensions  \cite{g1,g2}
can be realized by a single
fivebrane of M theory wrapping a Riemann surface.
The fivebrane configurations corresponding to these $N=1$
supersymmetric gauge theories  encode the information about the
$N=1$ moduli spaces of vacua \cite{hoo,W,N1,N2,N3,N4,N5,N6,N7,N8,N9,N10}

So far the fivebrane construction has been successfully applied to
the computation of holomorphic (or rather BPS)
 quantities of the four dimensional
supersymmetric gauge theory. Of particular importance are the non-holomorphic
quantities such as higher derivative terms in $N=2$ theories and
the {\Kahler} potential of $N=1$ supersymmetric gauge theories.
It is very difficult to compute these objects by field theory
techniques in particular in regions of strong coupling.

There are two questions to be asked: Can we compute these non
holomorphic
quantities using the fivebrane and do the results agree
with what we expect for the gauge theories in four dimensions?
As is well known, the theory on the M theory fivebrane
is a $(0,2)$ theory in six dimensions.
When wrapping a Riemann surface $\Sigma$ the four dimensional theory
on ${\bf R}^4$ has two scales: The radius of the eleventh dimension $R$ and
the typical scale of the brane configuration $L_{brane}$.
There are two corresponding Kaluza-Klein modes with masses $1/R$ and
$1/L_{brane}$.
The four dimensional gauge field theories
that we are interested in have one scale $\Lambda$.
In order to correctly obtain these four dimensional
theories we have to find the region of values of the parameters
$R,L_{brane}$ where the wrapped fivebrane theory
and the gauge theory agree. This in particular requires a
decoupling of the Kaluza-Klein modes.

Holomorphic  quantities of field theory are not sensitive to the
region
of parameters $R,L_{brane}$ where they are computed. In particular
they can also be computed in the eleven dimensional
supergravity limit of M theory even though the $1/R$ Kaluza-Klein
modes are light.
It is not clear whether this holds also for the computation of
the non-holomorphic quantities.
The aim of this paper is to answer this question.
We will use the fivebrane of eleven dimensional
supergravity in order to compute non holomorphic quantities
and compare with what we expect from field theory.

The paper is organized as follows:

Section 2 includes a brief discussion of the eleven dimensional
supergravity fivebrane action which will be needed for the
computations.
It also contains
a derivation of the formula for the {\Kahler} potential of the
four dimensional theory obtained via wrapping the fivebrane on a
Riemann
surface.
In section 3 we use this formula
to
show that the {\Kahler} potential on the Coulomb branch of $N=2$
theories
is correctly reproduced.
In section 4 we compute the four derivative term in the
$N=2$ effective action which is a non holomorphic quantity.
We find an explicit dependence on the radius $R$ of the eleventh
dimension.
We compare the result
with what we expect for the four dimensional gauge field theory and
show that the results disagree for any value of $R$.
In section 5 we use the fivebrane
to compute the {\Kahler} potential of
$N=1$ gauge theories in an abelian Coulomb phase.
We compare the brane result with
what we expect for the four dimensional gauge field theory.
Although the effective coupling is correctly reproduced
the K\"ahler potential again disagrees with
what we expect for the field theory.
In section 6 we study the Higgs branch of $N=2$ theories.
In this case the fivebrane worldvolume consists
of several disjoint components whose motions parametrize the Higgs
branch. We compute the effect of the
gravitational force  on each component due to the other
components and obtain correctly the metric of the $N=2$
Higgs  branch, up to 
possible corrections due to
membranes wrapping supersymmetric
3-cycles.
In section 7 we study the metric on the 
$N=1$ Higgs branch using the same
method. We consider $N_f=N_c=2$ SQCD with and without
a heavy adjoint chiral multiplet.
In the case with heavy adjoint,
we find an indication that the complex structure 
is correctly reproduced and the result leads to a
proposal on the precise
relation between the flavor rotation
and the motion of a finite
component of the fivebrane.
In the case without adjoint, the computation
only captures classical features of the metric
on the Higgs branch.
Section 8 is devoted to a discussion of the results.

\section{Preliminaries}

At low energies, $M$ theory is described by eleven dimensional
supergravity, and the classical action describing the $M$ theory fivebrane 
has been determined in \cite{five1,five2,five3}. We will only consider
the bosonic part of the fivebrane action, the rest of the fivebrane
action is determined by supersymmetry. The world-volume fields
of the fivebrane consist of fields $X^{\mu}(x^a)$, where $\mu=0,\ldots,10$
and $a=0,\ldots,5$, describing the embedding of the six-dimensional
world-volume in eleven-dimensional spacetime, and of a self-dual
two-form $B_{ab}$. In addition, the fivebrane action will depend on
the eleven-dimensional background fields, which are the metric $G_{\mu\nu}$
and the three-form $C^{(3)}_{\mu\nu\rho}$. The bosonic part of the action
expanded up to second order in $B_{ab}$ reads, up to a Wess-Zumino term,
\be
\label{fiveact}
S\,\,=\,\, \frac{1}{\ell_{11}^6} \int \dd^6 x\,
\sqrt{-g}\, + \,\int \dd^6 x \,|dB-C^{(3)}|^2\,,
\ee
where $g=\det(g_{ab})$ and $g_{ab}$ is the induced metric
\be
\label{indmetric}
g_{ab} = G_{\mu\nu} \partial_a X^{\mu} \partial_b X^{\nu}.
\ee
The self-duality constraint for the two-form is that $dB-C^{(3)}$
should be a self-dual three-form with respect to the induced metric
$g_{ab}$. In fact, (\ref{fiveact}) does not completely define 
the theory of a self-dual three-form. For that, one has to add an
additional term involving an auxiliary scalar as in \cite{five1}
which yields in a specific gauge the formulation of \cite{five2}. 
Alternatively, one can extract from the partition function of (\ref{fiveact})
the piece relevant for the self-dual three form as in \cite{witten2}.

The theories that we consider in this paper are obtained from fivebranes
of the form ${\bf R}^4 \times \Sigma$ embedded into a spacetime of
the form ${\bf R}^4 \times M^7$. Here, the two ${\bf R}^4${}'s are to
be identified with each other, and $\Sigma$ is a Riemann surface
embedded in $M^7$. By performing a Kaluza-Klein reduction of the
fivebrane theory on the Riemann surface $\Sigma$ we obtain a four-dimensional
theory. As we discuss later, this reduction can be quite subtle,
especially in the case where the Riemann surface is non-compact.
The various bosonic fields in the four-dimensional theory 
arise as follows. First, in general there will be family of Riemann surfaces
$\Sigma(u_{\alpha})$ depending on moduli $u_{\alpha}$. These moduli become scalar
fields in the four-dimensional theory. Second, if the Riemann surface
has components of finite volume, additional scalar fields arise by
taking $B_{ab}$ proportional to the volume form of one of these components.
 Finally, if the Riemann surface has genus $g$ greater than zero there will
be $g$ $U(1)$ vector fields coming from the decomposition of $B_{ab}$ 
in terms of the harmonic one-forms on $\Sigma$. Although the full fivebrane
action is rather complicated, the two terms given in (\ref{fiveact}) will
be sufficient for our purposes, as these are the only ones contributing
to the terms involving two derivatives in the four dimensional  theory.
In the cases where we consider higher derivative terms $B$ and $C^{(3)}$
do not contribute, and (\ref{fiveact}) will again be all we need.

To determine the action of the four-dimensional theory, we need to 
consider fivebranes of the form ${\bf R}^4 \times \Sigma$ where $\Sigma$ is
allowed to vary over ${\bf R}^4$. More precisely, we assume that
the metric on spacetime ${\bf R}^4 \times M_7$  is of the form
\be
ds^2 = f(X^i) \eta_{mn} dX^m dX^n + G_{ij} dX^i dX^j
\ee
with $m=0,\ldots,3$ and $i=4,\ldots,10$, and consider fivebranes
with world-volume coordinates $x^m,z,\bar{z}$
whose embeddings are of the form
\bea
&& X^m = x^m, \qquad m=0,\ldots,3 \non \\{}
&& X^i = X^i(z,\bar{z},u_{\alpha}(x^m)), \qquad i=4,\ldots,10.
\label{embed}
\eea
Here, $z,\bar{z}$ are arbitrary coordinates on the Riemann
surface $\Sigma(u_{\alpha})$. As the fivebrane action is invariant
under world-volume diffeomorphisms, we can always choose 
$z$ and $\bar{z}$ in such a way that the induced metric on the
Riemann surface is conformal, i.e. $g_{zz}=g_{\bar{z}\bar{z}}=0$.
As this will simplify things considerably, we will from now
on always assume this to be the case.

The first term in the fivebrane action (\ref{fiveact}), when evaluated
for (\ref{embed}), yields
\be \label{5act}
S=\frac{1}{\ell_{11}^6} \int d^4 x \, d^2 z \, g_{z\bar{z}}\, \sqrt{-\det(f(X^i)
\eta_{mn} + L_{mn})}
\ee
where
\be
L_{mn} =  \frac{\partial u_{\alpha}}{\partial x^m}
\frac{\partial u_{\beta}}{\partial x^n}
 \left( g_{\alpha\beta}  -
 g_{\alpha z} \frac{1}{g_{z\bar{z}}}
 g_{\beta \bar{z}} -  g_{\beta z} \frac{1}{g_{z\bar{z}}}
 g_{\alpha \bar{z}} \right)
\ee
and 
\be 
\label{othermetric}
g_{\alpha \beta}=\frac{\partial X^i}{\partial u_{\alpha}} 
 G_{ij}  \frac{\partial X^j}{\partial u_{\beta}}, \qquad
g_{{\alpha}z} = 
 \frac{\partial X^i}{\partial u_{\alpha}} 
 G_{ij} \frac{\partial X^j}{\partial z}.
\ee
In particular, the kinetic term for the scalars $u_{\alpha}$ reads
\be \label{skin}
S_{kin}=\frac{1}{\ell_{11}^6} \int d^4 x \, d^2 z\, g_{z\bar{z}}\, f(X^i)
 \Tr(L).
\ee 

A simplification arises when spacetime is of the form ${\bf R}^4\times M_6 \times
 {\bf R}$, where $M_6$ is a K\"ahler manifold, and when $\Sigma$ is
holomorphically embedded in $M_6$ and depends holomorphically on the
moduli $u_{\alpha}$ (which should therefore be complex). 
We also take $f(X^i)=1$, so that the metric reads
\be
ds^2 = \eta_{mn} dX^m dX^n + 2 G_{i\bar{j}} dX^i dX^{\bar{j}} +
 (dX^{10})^2.
\ee
In this case the fivebrane
configuration preserves $N=1$ supersymmetry in four dimensions and
the kinetic term for the scalars should be given by a K\"ahler metric.
The kinetic term is given by
\be \label{finkin}
S= \int d^4 x\, \partial_m u_{\alpha} \partial^m u_{\bar{\beta}}\,
 K_{\alpha \bar{\beta}}
\ee
with
\be \label{finmet}
K_{\alpha \bar{\beta}} = \frac{2}{\ell_{11}^6} \int_{\Sigma} d^2 z (g_{z\bar{z}}
 g_{\alpha \bar{\beta}} - g_{z\bar{\beta}} g_{\alpha \bar{z}} ).
\ee 
This metric is indeed a K\"ahler metric. In fact, it is easy to give
 an expression for the corresponding K\"ahler potential. If 
$K_{{\rm spacetime}}(X^i,X^{\bar{i}})$ is the K\"ahler potential
for the spacetime metric $G_{i\bar{j}}$, i.e. $G_{i\bar{j}}= 
\partial_i \partial_{\bar{j}} K_{{\rm spacetime}}$, then
\be
K_{{\rm field theory}}=\frac{1}{2\ell_{11}^6} \int_{\Sigma} d^2 z
 \, g_{z\bar{z}} \, K_{{\rm
spacetime}}(X^i(z,u_{\alpha}),X^{\bar{i}}(\bar{z},u_{\bar{\alpha}})).
\ee
In words, the K\"ahler potential of the $N=1$ field theory in four
dimensions is the integral of the spacetime K\"ahler potential
over $\Sigma$ with its induced metric. 

The second term in (\ref{fiveact}) can also contribute to the kinetic
terms for the scalars and gauge fields in four dimensions. If the
Riemann surface has genus $g$, the decomposition of $B_{ab}$ in terms of
the harmonic one-forms on the Riemann surface yields $g$ $U(1)$
vector fields in four dimensions. The gauge coupling for these vector
fields is given by the imaginary part of the period matrix of the
Riemann surface, $S\sim \int ({\rm Im}(\tau_{ij})F_i \wedge \ast F_j)$
\cite{witten1}. The contribution of the second term in (\ref{fiveact})
to the kinetic term of the scalars is more complicated and depends on
the precise situation. We will only make a few general remarks and
postpone a more detailed discussion until we meet concrete examples
where this second term is relevant. 

The second term in (\ref{fiveact}) is invariant under $\delta
B=C^{(2)}$, $\delta C^{(3)}= dC^{(2)}$. The two-form $B$ can
give rise to additional scalar and vector fields in four dimensions.
Given some background three-from $C^{(3)}$, we should according
to the principle of Kaluza-Klein reduction choose $B$ 
in such a way that $\int |d B - C^{(3)}|^2$ gives
rise to kinetic terms for the four-dimensional fields only, without
a mass term, as we are only interested in the four-dimensional
fields that are massless. In addition, we have to worry about
the gauge invariance and the self-duality condition. Thus, in determining
$B$ we should impose two additional constraints. First, $dB-C^{(3)}$ should
be a self-dual three-form, and second, certain gauge fixing conditions
should be satisfied. As gauge fixing conditions we will use the Lorentz
gauges $d\ast(B-B_0)$ and $d\ast(C^{(3)}-C_0^{(3)} )=0$, for some
fixed $B_0$ and $C_0^{(3)}$. 

A special situation is when 
the pull-back of $dC^{(3)}$ to the fivebrane world-volume vanishes, in which 
case there exists a two-form $C^{(2)}$ on the world-volume such that
$C^{(3)}=dC^{(2)}$, and we can take $B=C^{(2)}+B'$. The second
term in (\ref{fiveact}) then simply reads $\int |dB'|^2$.
The only cases we will encounter where the pull-back of $dC^{(3)}$
is non-vanishing is when we consider one fivebrane in the background
of another five-brane. A five-brane induces a background geometry
with a non-trivial $dC^{(3)}$ which is, roughly speaking, the unit
volume form on the four-spheres surrounding the five-brane in
the five dimensions transversal to it \cite{guv}. When this happens, the
kinetic terms for the scalars $u_{\alpha}$ will also be modified
because $C^{(3)}_{mz\bar{z}}$ contains terms $\partial_m u_{\alpha}
\partial_{\alpha} X^i \partial_z X^j \partial_{\bar{z}} X^k
C^{(3)}_{ijk}$. 

A final observation is one regarding the additional scalar fields
coming from $B$. As we mentioned previously, if the Riemann surface $\Sigma$
has a component $\Sigma_0$ of finite volume we get
an additional scalar $\sigma$ by
taking $B_{z\bar{z}}=\sigma(x^m) \omega_{z\bar{z}}$, with
$\omega_{z\bar{z}}$ the volume form on $\Sigma_0$. It turns out that
$\sigma$ is a compact scalar. To see this, we replace ${\bf R}^4$ by
${\bf R}^3 \times S^1$. According to \cite{witten2}, in order to
be able to define the five-brane partition function, we do not
only have to mod out by the gauge transformations $\delta B=C^{(2)}$,
$\delta C^{(3)} = dC^{(2)}$, but also by `large' gauge transformations
where we add an element of 
$H^3({\bf R}^3\times S^1 \times \Sigma,{\bf Z})$ to both $dB$ and
$C^{(3)}$. These large gauge transformations show that we
should identify $\sigma$
with $\sigma+{\rm const}$.

\section{$N=2$ Coulomb Branch}

In \cite{witten1}, Witten constructed configurations of the \MT fivebrane
starting from the configurations of D4, NS5 and D6 branes
in Type IIA string theory
that describe at long distances the dynamics of
$N=2$ supersymmetric gauge theories in four dimensions.
The Type IIA configuration is in a flat ten-dimensional space-time
with time and space coordinates $x^0$ and $x^1,\ldots,x^9$,
and the D4, NS5 and D6 branes span the directions 01236, 012345
and 0123789 respectively.

The \MT configuration for the $SU(N_c)$ gauge theory
with $N_f$ fundamental hypermultiplets ($N=2$ SQCD)
is a fivebrane embedded
in the eleven-dimensional space-time $\R^7\times S$ where $S$ is the
$A_{N_f-1}$-type Taub-NUT space which
is a four-dimensional non-compact hyper-K\"ahler manifold.
The $\R^7$ spans the 0123789 directions, while $S$ spans
the 456 directions in the Type IIA limit and wraps on the circle
in the eleventh direction.
Choosing one of its complex structures, $S$ is
described by a resolution of the complex surface
\beq
xy=\Lambda^{2N_c-N_f}\prod_{i=1}^{N_f}(v+m_i),
\eeq
and is provided with the nowhere vanishing holomorphic two-form
\beq
\Omega=\elel^3\, \dd v\wedge {\dd y\over y}\,.
\eeq
The parameters $m_i$ and the parameters of the resolution
(the size of the resulting two-spheres)
determine the location of the D6 branes in the 45 and 6
directions respectively, where $m_i$'s correspond
to the quark bare mass but the parameters of the resolution
have no counterpart in the standard gauge theory.
$\Lambda$ is a parameter corresponding to the dynamical scale.

The fivebrane for a theory at a point in the Coulomb branch
is $\R^4\times \Sigma$ where $\R^4$ spans the first four directions
0123 while $\Sigma$ is at a point in
the last three directions 789 of $\R^7$ and
is embedded as a holomorphic curve in $S$.
The embedding is given by
\beq
x+y=2C_{N_c}(v,u_{\alpha}):=
2\left(\,v^{N_c}+\sum_{\alpha=2}^{N_c}u_{\alpha}v^{N_c-\alpha}
\,\right)\,.
\eeq
This is the same as the Seiberg-Witten curve \cite{sw1,klty,af,us,apsp}
of $N=2$ SQCD and, therefore, the worldvolume theory has the
same effective gauge coupling as $N=2$ SQCD \cite{Verlinde}.
A BPS state is a supersymmetric membrane
ending on the fivebrane worldvolume \cite{KLMVW,witten1}
whose mass is given by the membrane tension $\elel^{-3}$ times
the area of the spacial part $D$ of its worldvolume.
By the condition of the supersymmetric cycle,
the area is the same as the integration of the holomorphic two-form
$\Omega$ and the mass is given by
\beq
{1\over \elel^3}\left|\,\int_D\Omega\,\right|
=\left|\,\int_D\dd v\wedge {\dd y\over y}\,\right|
=\left|\,\oint_{\partial D} v{\dd y\over y}\,\right|\,,
\eeq
which agrees with the BPS mass formula for the $N=2$ SQCD,
as shown in \cite{Fayy,Piljin,Mikhailov} for the $N_f=0$ cases.

 From the fact that the worldvolume theory has the same effective
gauge coupling as $N=2$ SQCD, it follows that they
have the same K\"ahler metric on the Coulomb branch as well
because of the $N=2$ supersymmetry
in four dimensions which both theories possess.
We now explicitly check this.
\footnote{Essentially the same computation was recently done
independently in \cite{Mikhailov,hlw3}.}

Since we are considering a variation of the holomorphic curve $\Sigma$
with a single non-compact component,
the K\"ahler metric is simply read off
from the second order variation of
the term $\elel^{-6}\int\sqrt{-g}\dd^6x$ of the fivebrane lagrangian.
Recalling that $\Sigma$ is at a point in the 789 directions
and is embedded in the complex surface $S$, we find that
it is given by formula (\ref{finmet}). Namely
\beq
K_{\alpha\bar \beta}
={1\over \elel^6}\int_{\Sigma}\dd^2z(G_{i\bar j}G_{k\bar l}-
G_{i\bar l}G_{k\bar j})\partial_zX^i\partial_{\bar z}X^{\bar j}
{\partial X^k\over \partial u_{\alpha}}
{\partial X^{\bar l}\over \partial \overline{u}_{\bar \beta}}\,,
\label{FINMET}
\eeq
where $G_{i\bar j}$ denotes the K\"ahler metric of $S$.  
Note that $G_{i\bar j}G_{k\bar l}-
G_{i\bar l}G_{k\bar j}$ is the $i\bar j k\bar l$ component of
the square of the K\"ahler form $\omega$ of $S$.
In the present case
where $S$ is a Ricci-flat K\"ahler manifold of dimension two,
the square of the K\"ahler form is given by
\beq
\omega^2=\Omega\wedge\overline{\Omega}\,.
\label{tt*}
\eeq
We now fix the coordinates of the space-time $S$ and the worldvolume
$\Sigma$. As the coordinates of $S$ we can use $v$ and $y$ which are
good in the neighborhood of $\Sigma$ except at a subset of measure
zero. As the worldvolume coordinate, we choose $v$. This choice
of coordinates corresponds to considering the curve $\Sigma$
as the two-sheeted cover of the $v$-plane given by 
\beq
y^2-2C_{N_c}(v,u_{\alpha})y
+\Lambda^{2N_c-N_f}\prod_{i=1}^{N_f}(v+m_i)=0\,.
\eeq
Then, there is only one non-trivial component of
$\partial X^k/\partial u_{\alpha}$ which is
\beq
{\partial y\over \partial u_{\alpha}}
=y{v^{N_c-\alpha}\over y-C_{N_c}(v)}\,.
\eeq
Note that
\beq
\omega_{\alpha}={v^{N_c-\alpha}\over y-C_{N_c}(v)}\dd v\,,\qquad
\alpha=2,\ldots,N_c
\eeq
form a base of the holomorphic differentials of the curve $\Sigma$
of genus $N_c-1$.
The K\"ahler metric is then expressed as
\beqa
K_{\alpha\bar \beta}&=&
{1\over \elel^6}\int_{\Sigma}\dd^2v\,\Omega_{vy}
\overline{\Omega}_{\bar v\bar y}
{\partial y\over \partial u_{\alpha}}
{\partial \bar y\over \partial \bar u_{\bar \beta}}
=\int_{\Sigma}\dd^2v\,
{v^{N_c-\alpha}\over y-C_{N_c}(v)}
\overline{\left({v^{N_c-\bar \beta}\over y-C_{N_c}(v)}\right)}
\nonumber\\
&=&\int_{\Sigma}\omega_{\alpha}\wedge\overline{\omega}_{\bar \beta}
\label{K1}
\eeqa
This is nothing but the K\"ahler metric of the special geometry
which agrees with the field theory knowledge.
In order to see this in the standard notation,
let us choose a symplectic basis of the first homology class
of $\Sigma$; $A_i,B^j$ ($i,j=1,\ldots, N_c-1$).
Then, the $a$ and $a_D$ fields are given by
$\partial a_i/\partial u_{\alpha}=\oint_{A_i}\omega_{\alpha}$
and
$\partial a_D^j/\partial u_{\beta}=\oint_{B^j}\omega_{\beta}$.
Using the Riemann bilinear identity, (\ref{K1}) is expressed
as
\beq
K_{\alpha\bar \beta}
=\sum_{i=1}^{N_c-1}\left({\partial a_i\over \partial u_{\alpha}}
{\partial \overline{a_D^i}\over \partial \overline{u}_{\bar \beta}}
-{\partial \overline{a_i}\over \partial \overline{u}_{\bar \beta}}
{\partial a_D^i\over \partial u_{\alpha}}\right)\,.
\eeq
In this way, we have obtained the standard form of the
scalar kinetic term of the effective lagrangian
\beq
S_{\rm kin}={\rm Im}\int\dd^4x\,\eta^{mn}
\sum_{i=1}^{N_c-1}\partial_ma_i\partial_n\overline{a_D^i}\,.
\eeq

Note that the essential point for obtaining the special geometry
is the holomorphic - anti-holomorphic
factorization (\ref{tt*}) of the square of the K\"ahler form. 
This would not be the case
if $\Sigma$ were embedded in a Calabi-Yau three-fold (as in the
case of $N=1$ SQCD)
nor in a Ricci-non-flat complex surface in the space-time
(as in the case we will consider in section 5).

\section{$N=2$ Higher Derivative Terms}

In this section we will consider the four-derivative terms for the scalar fields
in $N=2$ SYM theory, both from the field theory point of view and from the
brane point of view, and compare the results.

\subsection{Field Theory}

The low-energy effective action of $N=2$ SYM theory can in $N=2$ superspace
be written as
\be \label{4deract}
S= \int d^4 x d^4 \theta F(A^i) + \int d^4 x d^4 \bar{\theta} \bar{F}
(\bar{A}^i) + \int d^4 x d^4 \theta d^4 \bar{\theta}
H(A^i,\bar{A}^{\bar{i}})+ \ldots,
\ee
where $F$ is a holomorphic function and $A^i$ ($\bar{A}^{\bar{i}}$)
are abelian $N=2$ chiral (antichiral) vector superfields. The real
function $H(A^i,\bar{A}^{\bar{i}})$ is the one that gives rise to
four derivative terms for the scalars in the low-energy effective
action. We denote by $\Phi^i$ and $W^i_{\alpha}$ the $N=1$ chiral
superfield and $N=1$ field strength that are contained in $A^i$, and
by $\phi^i$ the complex scalar in $\Phi^i$. The kinetic term for
the $\phi^i$ is
\be
S=\int d^4 x \, (\partial_m \phi^i \partial^m \phi^{\bar{j}} )
K_{i\bar{j}}
\ee
where $K_{i\bar{j}}\sim {\rm Im}(F_{ij}(\phi^i))$ (subscripts on 
$F$ and $H$ denote derivatives with respect to $\phi^i$).  

The function $H$ been studied in 
\cite{henn,mat,yung,ketov,wgr,lgru,bfmt,bko}. Some general facts such as
its behavior under $SL(2,{\bf Z})$ were discussed in \cite{henn}, and
its asymptotic behavior was discussed in \cite{wgr,lgru}. Several
contributions to $H$ are known explicitly, such as the one-loop
contribution \cite{ketov}, 
the two-loop contribution \cite{bko} (which vanishes),
the one-instanton contribution \cite{yung}
and the two-instanton contribution \cite{bfmt}. An exact form
for $H$ in the case of $SU(2)$ was conjectured in \cite{mat}. 
In terms of the gauge invariant coordinate $u$ on the Coulomb branch
of the $SU(2)$ gauge theory it states that
\be
H(u,\bar{u})= c |u^2 - \Lambda_{N=2}^4 | (K_{u\bar{u}})^2,
\ee
where $c$ is some constant. $H(u,\bar{u})$ should really be seen as
a function of $A,\bar{A}$ rather than $u,\bar{u}$, 
because a holomorphic function of an $N=2$
vector superfield is in general no longer an $N=2$ vector superfield. 
Such a function is still chiral, but no longer satisfies the
Bianchi identity. 
The one-loop results for $SU(2)$ are  
\be
\label{oneloop}
 K_{u\bar{u}}\sim \frac{ \log(16 u\bar{u} /
\Lambda_{N=2}^4)}{\sqrt{u\bar{u}}},
\qquad u= A^2/2, \qquad 
H(A,\bar{A}) \sim \log(A/\Lambda_{N=2}) \log(\bar{A}/\Lambda_{N=2}). 
\ee
In addition, there are instanton correction that are typically
weighted with factors $(\Lambda_{N=2}^4/u^2)$, but we will not
consider those here.

As the brane will not give us an answer in superspace but in
components, in order to be able to compare we need to work out
what (\ref{4deract}) looks like in components. Using the $N=1$
expansion of (\ref{4deract}) in e.g. \cite{henn} or \cite{wgr},
we find the following four-derivative terms for the scalars
$\phi^i$
\bea
S_{4} & = & \int d^4 x ( 
 2 H_{i\bar{j}} (\partial^m \partial_m \phi^i) (\partial^n \partial_n
\phi^{\bar j}) + 
H_{ij\bar{k}}  (\partial^m \phi^i)  (\partial_m \phi^j) (\partial^n \partial_n
\phi^{\bar k})
\non \\{}
& & \quad + H_{\bar{i}\bar{j}{k}}
  (\partial^m \phi^{\bar i})  (\partial_m \phi^{\bar j}) (\partial^n \partial_n
\phi^{k})
+ H_{ij\bar{k}\bar{l}}  (\partial^m \phi^i)  (\partial_m \phi^j) 
 (\partial^n \phi^{\bar k})  (\partial_n \phi^{\bar l}) ).
\label{4dercomp}
\eea
In deriving this result we used several partial integrations, and put
the auxiliary fields equal to zero. If the fermions are equal to zero that
is allowed because there are no terms linear in the auxiliary fields
that couple only to the scalar fields $\phi^i$. This can easily be
understood  from the fact that the scalars $\phi^i$ are invariant under the global
$SU(2)_R$ transformations, whereas the three auxiliary fields form a 
triplet. We can simplify the structure of the four-derivative term
even further by making some field redefinitions of the $\phi^i$. The
field equation for $\phi^i$ reads
\be \label{aux1}
\partial^m \partial_m \phi^i = - (K_{i\bar{j}})^{-1} 
K_{\bar{j} kl} (\partial^m \phi^k) (\partial_m \phi^l),
\ee
and using field redefinitions we can replace $\partial^m \partial_m
\phi^i$ in (\ref{4dercomp}) by the right hand side of (\ref{aux1}).
This then finally leads to the following expression for the
four-derivative term
\be \label{4dfin1}
S_{4} = \int d^4 x (\tilde{H}_{ij\bar{k}\bar{l}} 
 (\partial^m \phi^i)  (\partial_m \phi^j) 
 (\partial^n \phi^{\bar k})  (\partial_n \phi^{\bar l})  )
\ee
where
\be \label{4dfin2}
 \tilde{H}_{ij\bar{k}\bar{l}} =H_{ij\bar{k}\bar{l}}
- H_{ij\bar{p}} (K_{\bar{p}q})^{-1} K_{q\bar{k}\bar{l}}
- K_{ij\bar{p}} (K_{\bar{p}q})^{-1} H_{q\bar{k}\bar{l}}
+2 K_{ij\bar{p}} (K_{\bar{p}q})^{-1} H_{q\bar{r}}
 (K_{\bar{r}s})^{-1} K_{s\bar{k}\bar{l}}.
\ee
It is quite interesting that the four-derivative terms can be brought
in the simple form (\ref{4dfin1}), a fact that the brane knows about 
as we will see in the next section. For $SU(2)$, we find that the
semi-classical four-derivative term reads
\be \label{su24d}
S=\int d^4 x (\partial^m u \partial_m u)  (\partial^n \bar{u}
\partial_n {\bar u})
\frac{8+4 \log y + (\log y)^2}{u^2 \bar{u}^2 (\log y)^2}, \qquad
y=\frac{16 u \bar{u}}{\Lambda_{N=2}^4}.
\ee

\subsection{Fivebrane}

Let us now compute the higher derivative terms using the
fivebrane\footnote{These terms were also recently computed in
\cite{hlw3}}. They can be extracted by expanding (\ref{5act}) in powers of
$L$. We will concentrate on the case of a pure gauge theory without
matter, with target space metric as given in (\ref{11d}).
Recall that $\Sigma$ is described by
\be t^2 - 2 t C(v,u_{\alpha}) + \Lambda_{N=2}^{2N_c} =0, 
\qquad C(v,u_{\alpha}) = v^{N_c} + \sum_{\alpha=2}^{N_c} u_{\alpha} v^{N_c-\alpha}.
\ee
We identify the complex coordinate $z$ on $\Sigma$ with $v$ and find 
(where $C\equiv C(v,u_{\alpha})$ and $C' \equiv \partial
C(v,u_{\alpha}) /\partial v$)
\bea
g_{z\bar{z}} & = & \frac{\ell_{st}^4 |t-C|^2 + R^2 |C'|^2 }{|t-C|^2}
 \non \\{}
g_{\alpha \bar{\beta}} & =  & \frac{R^2 v^{N_c-\alpha} {\bar
v}^{N_c-\bar{\beta}}}{|t-C|^2}
 \non \\{}
g_{z\bar{\beta}} & = & \frac{R^2 C' \bar{z}^{N_c-\bar{\beta}}
}{|t-C|^2}
\non \\{}
g_{\alpha\bar{z}} & = & \frac{R^2 \bar{C'} z^{N_c-\alpha} }{|t-C|^2}.
\label{metcomp}
\eea
 From this we obtain
\be
L_{mn} = (\partial_m u^{\alpha} \partial_n {\bar u}^{\bar{\beta}} ) 
\frac{ \ell_{11}^6 v^{N_c-\alpha} {\bar v}^{N_c-\bar{\beta}} }{
 \ell_{st}^4 |t-C|^2 + R^2 |C'|^2} + (m \leftrightarrow n).
\ee
The five-brane action (\ref{5act}) can now be rewritten in the
following convenient form: introduce $f,d$ defined by
\bea
f & = & 
 (\partial^m u^{\alpha} \partial_m u^{{\beta}} ) 
\frac{ \ell_{11}^6 v^{2N_c-\alpha-\beta} }{
 \ell_{st}^4 |t-C|^2 + R^2 |C'|^2} \non \\{}
d & = & 
 (\partial^m u^{\alpha} \partial_m \bar{u}^{\bar{\beta}} ) 
\frac{ \ell_{11}^6 v^{N_c-\alpha} {\bar v}^{N_c-\bar{\beta}} }{
 \ell_{st}^4 |t-C|^2 + R^2 |C'|^2},
\eea
then
\be
S=\frac{1}{\ell_{11}^6} \int d^4 x d^2 v 
\frac{\ell_{st}^4 |t-C|^2 + R^2 |C'|^2 }{|t-C|^2}
\sqrt{(1+d)^2 - f\bar{f}}.
\label{aux2}
\ee
This expression makes it clear what kind of higher derivative terms
can appear in the fivebrane action, and what their relative
coefficients are. In particular, the four-derivative term comes only
from the $f\bar{f}$ term in (\ref{aux2}), and it is therefore
precisely of the same form as (\ref{4dfin1}), namely
\be \label{aux6a}
S_4 = \int d^4 x \tilde{H}_{\alpha\beta\bar{\gamma}\bar{\delta}} 
(\partial^m u_{\alpha} \partial_m u_{\beta}) 
(\partial^n \bar{u}_{\bar{\gamma}} \partial_n \bar{u}_{\bar{\delta}})
\ee
with
\be \label{aux6}
\tilde{H}_{\alpha\beta\bar{\gamma}\bar{\delta}} = 
-\frac{1}{2} \int_{\Sigma} d^2 v \frac{\ell_{11}^6
v^{2N_c-\alpha-\beta}
 {\bar v}^{2N_c-\bar{\gamma}-\bar{\delta}} }{ |t-C|^2 ( \ell_{st}^4
|t-C|^2 + R^2 |C'|^2 )}.
\ee
We see that the fivebrane provides a form of the four-derivative term
that is consistent with the general form of a four-derivative term in
an $N=2$ field theory, provided we make certain field
redefinitions. In itself this is not surprising, as the fivebrane
theory does have $N=2$ supersymmetry in four dimensions, and should
therefore be consistent with the most general supersymmetric field
theory. The fact that we have to perform certain field redefinitions is 
presumably related to the fact that the $N=2$ supersymmetry in field theory is
realized differently than it is in the fivebrane theory. The latter
is known to be realized in a highly non-linear way \cite{hlw}. 
The fact that in the fivebrane theory fewer higher derivative terms
appear than in the field theory, suggests that the non-linear
realization of supersymmetry on the fivebrane is a very efficient
way to organize the higher derivative terms, and perhaps also the
fermionic terms, in $N=2$ effective actions. 

\subsection{Comparison}

The first thing that one notices in the brane result (\ref{aux6}) is
the fact that it depends non-trivially on $R$ and $\ell_{11}$, and
that it therefore cannot be equivalent to the field theory answer. 
This is the first time in this paper that we compare a field theory
quantity that is not protected by holomorphy or global symmetries to
the same quantity obtained from the brane. The brane result depends
explicitly on $R$ and $\ell_{11}$ and there is no obvious limit
for these quantities that yields a sensible result. The
four-derivative term is ill-defined both in the limit that $R$ goes
to zero and that $R$ goes to infinity. Therefore, the brane and field theory
results agree only for processes involving very low energies and
momenta. Certain qualitative features of the higher derivative terms
do agree. Consider for instance the points in moduli space
where (\ref{aux6}) becomes singular. This can only happen when
the denominator in the integrand behaves like $|v-v_0|^k$ near
some point $v=v_0$ and $k\leq 2$. There are two possibilities,
either $(t-C)^2$ has a double zero or $(t-C)^2$ and $C'$ have 
a common zero. The second condition implies the first one,
and $(t-C)^2$ has a double zero only at the singularities in the
moduli space where a dyon becomes massless. This is consistent
with what one expects from field theory.
It is in this sense that one might consider the
theories to be in the same universality class. 
To get an idea to what extent the results are quantitatively different, we compare
the semiclassical field theory result (\ref{su24d}) with the brane result for
$SU(2)$, which reads (see (\ref{aux6}))
\be
\label{aux}
\tilde{H}_{uu\bar{u}\bar{u}} \sim \int d^2 v
 \frac{\ell_{11}^6 }{ |(v^2 + u)^2 -\Lambda_{N=2}^4| ( \ell_{st}^4
|(v^2+u)^2 -\Lambda_{N=2}^4 | + 4 R^2 |v|^2 )}.
\ee
The semiclassical region corresponds to large $u$. If $|u| \gg R^2
 \ell_{st}^{-4}$ and $|u| \gg \Lambda_{N=2}^2$, then one can easily
 show that
\be |\tilde{H}_{uu\bar{u}\bar{u}} | \geq \frac{1}{4} \int_{|v| \geq \ell_{st}^2
 R^{-1} |u|} d^2 v \frac{R^2}{|v|^8} \sim R^8 \ell_{st}^{-12}
 |u|^{-6}.
\ee
 For these values of $|u|$ the integral over the disc $|v| \leq R
 \ell_{st}^{-2}$ contributes something of the order $R^4
 \ell_{st}^{-4} |u|^{-4}$ to $ |\tilde{H}_{uu\bar{u}\bar{u}} |$.
Altogether it is not clear whether (\ref{aux6}) does 
have the right $|u|^{-4}$ behavior, but if it does, the
coefficient in front of $|u|^{-4}$ will almost certainly 
depend on $R/\ell_{st}$. Thus, even in the semiclassical regime,
the field and brane theories are quantitatively
quite different. 

\subsection{$SU(N_c)$ with $N_f>0$}

In the presence of matter fields, field theory makes a few interesting
non-trivial predictions. The one-loop
contribution to $H(A^i,\bar{A}^i)$ has a coefficient proportional
to $2N_f-N_c$. In addition, 
there are indications \cite{ketov2,bko} that
$H(A^i,\bar{A}^i)$ receives no perturbative
corrections beyond one-loop at all.
Therefore, in the case where $N_c=2N_f$, 
the function $H(A^i,\bar{A}^i)$ should
receive contributions only from instantons.

Another prediction form field theory is that in the finite, scale 
invariant case $N_f=2N_c$, the one-loop result for $H(A^i,\bar{A}^i)$
is exact \cite{dise}. 

To see whether we can reproduce these results from the fivebrane, we
have to work out the four-derivative term in the presence of matter.
Matter is included in the $M$ theory framework by modifying space-time
to include a multi-Taub-NUT space. The Taub-NUT space depends on several
parameters $(x^4_i,x^5_i,x^6_i)$, corresponding to the locations of
the D6 branes in the Type IIA picture (see Appendix~B). Two of the
three parameters, $x^4_i$ and $x^5_i$, correspond to the mass of
a quark and should be taken equal to zero, but the other parameters
$x^6_i$ are free. They do not correspond to parameters in field theory.
If we work out the four-derivative term using the Taub-NUT background
we find that the result does depend on $x^6_i$. This illustrates once
more the difference between the brane results and field theory. A limit
that is particularly easy to analyze is to send all $x^6_i \rightarrow
\infty$. From the Type IIA point of view, this corresponds to using
semi-infinite four-branes to realize the matter. The result one 
obtains is given by (\ref{aux6a}) and (\ref{aux6}), where in 
(\ref{aux6}) $|C'|^2$ should be replaced by $|\frac{t-C}{t}
\frac{\partial t}{\partial v}|^2$. This result deviates considerably
from the field theory predictions. For $N_c=2N_f$ there are still
one-loop contributions to $H(A^i,\bar{A}^i)$, and for $N_f=2N_c$ 
the one-loop result is not exact. In the latter case, the field theory
result that the one-loop result is exact was obtained in \cite{dise} 
using global symmetries and a scaling argument. The reason that the
same argument does not apply to the brane calculation is that the 
fivebrane depends on two additional parameters, $R$ and $\ell_{st}$.
To have the same scale invariance as in \cite{dise} we should assign
weight $-1$ to both $R$ and $\ell_{st}$, and there are many new
scale invariant four-derivative terms that one can write down that
do depend on $R$ and $\ell_{st}$.

\section{$N=1$ Coulomb Branch}

In this section, we consider an example of a worldvolume theory
with $N=1$ supersymmetry in four-dimensions where we can obtain the
K\"ahler metric of the moduli space exactly within
the eleven-dimensional supergravity approximation of \MT.

The example we consider corresponds to
the $N=1$ supersymmetric gauge theory
with gauge group $SU(N_c)$ with a massless adjoint chiral multiplet
$\Phi$ and $N_f$ quark multiplets $Q^i$, $\tilQ_i$ (fundamental
and anti-fundamental chiral multiplets) with
bare mass $m$.
The classical Lagrangian of the theory is the standard D-term plus the
 F-term given by the superpotential
\beq
W=m\sum_{i=1}^{N_f}\tilQ_i Q^i\,.
\eeq
We do not turn on the Yukawa coupling $\tilQ\Phi Q$ which
makes the system $N=2$ supersymmetric.
The classical moduli space of vacua consists of a single
Coulomb branch where the quark VEVs are zero $Q=\tilQ=0$
and $\Phi$ is diagonal.
At energies below $m$ the field content of the theory is the same as
that of $N=2$ super-Yang-Mills theory with gauge group $SU(N_c)$,
but the quark multiplets introduce $N=2$ breaking interactions
and we do not expect to obtain the special geometry as the
quantum moduli space of vacua.
If the bare mass $m$ is very large, however, the $N=2$ breaking
interactions are suppressed as inverse powers of $m$,
and we do expect the moduli space to be
a small deformation of the special geometry for
the $N=2$ super-Yang-Mills theory with gauge group $SU(N_c)$.
We compute the corresponding deformation in the worldvolume theory,
and compare with what we expect from field theory.

There is no \apriori\ reason to expect that these two agree
because the supergravity computation is valid
when all the characteristic lengths of the
space-time and the brane are much larger than the Planck
length while the worldvolume theory becomes close to the
four-dimensional gauge theory only in the Type IIA limit
where the radius of the circle in the eleventh direction
is much smaller than $\elel$.
As in the $N=2$ Coulomb branch, the effective gauge coupling constant
will be correctly reproduced because
abelian gauge fields are obtained by the chiral two form on the
fivebrane whose lagrangian is scale invariant
and the result will persist to be true even if we scale the system
down to the Type IIA region $R\ll\elel$.
However, because of the lack of $N=2$ supersymmetry
it does not mean in the present case that
the K\"ahler metric is also correctly reproduced.

As we will see, there is indeed a clear discrepancy between the
two. The reason we dare to carry out this computation is
that the result might be useful for the following two purposes.
In the region $R\ll\elel$ the discrepancy would mainly be due to
the correction to the supergravity approximation of \MT
which is still in rather mysterious, and it can be used to learn about
\MT itself.
In the region $R\gg\elel$, the discrepancy originates in the effects of
the degrees of freedom which are not part of the four-dimensional gauge
theory. If we want to learn something about field theory from
the worldvolume theory of the brane in general,
it would be very useful to estimate the effects of such extra degrees of
freedom as accurately as possible. The present computation
may be considered as the first step towards obtaining some
quantitative information about them.

\subsection{Brane Construction}

We start by constructing a Type IIA brane configuration
whose worldvolume dynamics describes at long distances the 
$N=1$ gauge theory given above.
It involves $N_c$ D4 branes stretched between two NS 5-branes
with $N_f$ D6 branes located away from them.
The worldvolume of D4, NS5 and D6 branes span the directions
01236, 012389 and 0123789 respectively, where
the two NS 5-branes are at a point $v=0,x^7=0$ in the 457 directions
and are separated in the $x^6$ direction, the D4 branes stretched
between them are at points in the 89 directions,
and the D6 branes are at $v=-m,x^6=0$. Note that
the D4 and NS5 branes are separated from the D6 branes in the 45
directions by $\Delta v=m$.
This configuration is obtained from that of $N=2$ SQCD with
$N_f$ massive quarks by a 90 degree rotation of
the D4/NS5 system in the 45-89 directions while keeping
the D6 branes intact.
\begin{figure}[htb]
\begin{center}
\epsfxsize=2.5in\leavevmode\epsfbox{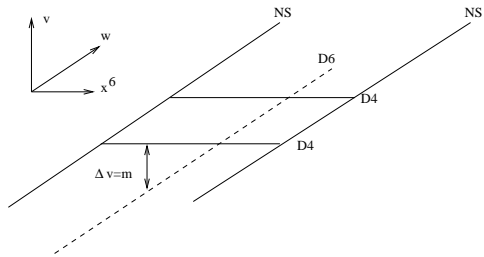}
\end{center}
\caption{The Type IIA Configuration}
\label{picture}
\end{figure}
The configuration is invariant under the groups $U(1)_{45}$
and $U(1)_{89}$ of rotations in the 45 and 89 directions
provided we assign a suitable $U(1)_{45}$ charge to $m$.
The coordinates $v$ and $x^8+ix^9$ carry charge $(2,0)$ and
$(0,2)$ respectively under $U(1)_{45}\times U(1)_{89}$,
and $m$ must be assigned a charge $(2,0)$.

The open strings stretched between the D4 branes create
a $U(N_c)$ vector multiplet in five dimensions with sixteen
supersymmetry, but the boundary condition at the ends on the
NS 5-branes projects this to an $N=2$ $SU(N_c)$ vector multiplet
in four dimensions which contains in $N=1$ language the vector
multiplet $W_{\alpha}$ and the adjoint chiral multiplet $\Phi$.
The rotation symmetry $U(1)_{45}\times U(1)_{89}$
is identified with the R-symmetry under which
$\Phi$ carries charge $(0,2)$
and gluino carries charge $(1,1)$.
Open strings stretched between D4 and D6 branes
create $N_f$ fundamental and anti-fundamental chiral multiplets
which carry $U(1)_{45}\times U(1)_{89}$ R-charge $(0,1)$.
As in the three-dimensional analog \cite{dhoy},
one can show that the superpotential invariant under the R-symmetry
is $W=m \tilQ Q$.

\newcommand{\w}{{\sl w}}
It is easy to lift this to an \MT configuration.
It is a configuration of a single fivebrane in the space-time
$\R^7\times S$ where $S$ is a Taub-NUT space
which has a complex structure (among others) described by
the equation
$
xy=\Lambda_{N=1}^{2N_c-N_f}(v+m)^{N_f}
$.
We have introduced in advance a parameter $\Lambda_{N=1}$
that characterize the distance between the two NS 5-brane.
The fivebrane is of the form $\R^4\times \Sigma$ where
$\Sigma$ is located at $x^7=0$ and is embedded as a holomorphic curve
in the space $S|_{v=0}\times \R^2$.
Here, $S|_{v=0}$ is the $v=0$ locus of $S$ which is the cylinder
described by
\beq
xy=m^{N_f}\Lambda_{N=1}^{2N_c-N_f}\,,
\eeq
while $\R^2$ is the 89 directions of the space-time in which
we introduce a complex coordinate\footnote{This 
should not be confused with the coordinate
$w=\elel^{-3}(x^8+ix^9)$ which is used in other parts of this paper.
The difference in the prefactor is because of the difference in
the identification of the parameters of branes and fields.}
\beq
\w=\elst^{-2}(x^8+ix^9)\,.
\eeq
The curve has two regions with large $\w$ corresponding to the two
NS 5-branes, and these obey the boundary condition 
$x\sim \w^{N_c}$ in one region of large $\w$ and
$y\sim \w^{N_c}$ in the other.
The embedding satisfying this condition
is given by
\beq
x+y=2C_{N_c}(\w,u_{\alpha})=2\left(\w^{N_c}
+\sum_{\alpha=2}^{N_c}u_{\alpha}\w^{N_c-\alpha}\right)\,,
\eeq
where $u_{\alpha}$'s are parameters characterizing the distance
between the $N_c$ D4-branes.

The configuration is invariant under $U(1)_{45}\times U(1)_{89}$
if we assign charge $(0,2N_c)$ to $x$ and $y$,
$(-2N_f,4N_c)$ to $\Lambda_{N=1}^{2N_c-N_f}$,
and $(0,2\alpha)$ to $u_{\alpha}$.
 From this we can identify $\Lambda_{N=1}$ and the $u_{\alpha}$'s
as the dynamical scale of the gauge theory and the color Casimirs
$\det(\w+\Phi)=C_{N_c}(\w,u_{\alpha})$ at some cut-off scale.
The effective gauge coupling constant of the
abelian gauge theory on the brane is given by the period matrix
of the curve.
The curve we have obtained reproduces the effective coupling 
of the gauge theory of interest given by \cite{Kapus}.

 For large values of $m$, the theory at energies below $m$
is approximately the $N=2$ pure Yang-Mills theory.
The dynamical scale of the low energy theory $\Lambda_{N=2}$
is related to the high energy theory by
\beq
\Lambda_{N=2}^{2N_c}=\left({g_L^2\over g_H^2}\right)^{N_c}
m^{N_f}\Lambda_{N=1}^{2N_c-N_f}\,,
\eeq
where $g_L$ and $g_H$ are the gauge coupling constants of the
low and high energy theories at the cut-off scale.
The factor $(g_L^2/g_H^2)^{N_c}$ appears \cite{NiMu}
since the kinetic term for $\Phi$ is given by
\beq
{1\over g^2}\int \dd^4 \theta {\rm Tr}(\Phi^{\dag}\Phi)\,,
\eeq
because the W-boson mass is given by the separation
of the eigenvalues of $\Phi$, not multiplied by $g$.
Likewise, the field $\Phi$ gets renormalized as we flow down 
the energy below $m$, and the low energy field $\Phi^{(2)}$
is related to the high energy field by
$\Phi^{(2)}=(g_L/g_H)\Phi$.
In terms of these variables, the curve is given by
\beq
\tilde{x}+\tilde{y}=2C_{N_c}(\tilde{\w},u^{(2)}_{\alpha})\,,
\qquad
\tilde{x}\tilde{y}=\Lambda_{N=2}^{2N_c}\,,
\eeq
where $u^{(2)}_{\alpha}=(g_L/g_H)^{\alpha}u_{\alpha}$
and $\tilde{x},\tilde{y},\tilde{\w}$ are suitably rescaled
coordinates.
This is nothing but the curve of $N=2$ super-YM theory,
and hence we have shown that the effective gauge coupling constant
is the same as the one in the $N=2$ super-Yang-Mills theory.
However, due to the absence of $N=2$ supersymmetry, the
scalar kinetic term (and hence the K\"ahler metric on the moduli
space) is different
as we now see.

\subsection{K\"ahler Metric on the Moduli Space}

\newcommand{\tilx}{\widetilde{x}}
\newcommand{\tilm}{\widetilde{m}}
\newcommand{\emst}{{M_{\it st}\!}}

As in the case of the $N=2$ Coulomb branch,
the K\"ahler metric on the moduli space is simply read off
from the second order variation of
the term $\elel^{-6}\int\sqrt{-g}\dd^6x$ of the fivebrane Lagrangian.
It is given by the formula (\ref{finmet}) or
(\ref{FINMET}) where
$G_{i\bar j}G_{k\bar l}-
G_{i\bar l}G_{k\bar j}$ is the $i\bar j k\bar l$ component of
the square of the K\"ahler form of $S|_{v=0}\times \R^2$.
By using the expression
of the Taub-NUT metric given in Appendix B, Equation
(\ref{TN2}), the K\"ahler form of $S|_{v=0}\times \R^2$ is
given by
\beq
\omega=R^2\,U^{-1}{\dd y\over y}\wedge{\dd\bar y\over\bar y}
+\elst^4\,\dd\w\wedge\dd \overline{\w}\,
\eeq
where
\beq
U=1+{N_f R\over 2\sqrt{|\elst^2 m|^2+(x^6)^2}}\,.
\eeq
The square of $\omega$ is thus
\beq
\omega^2=\elel^6\, U^{-1} \left|\dd \w\wedge{\dd y\over y}\right|^2\,.
\eeq
Proceeding as in the computation of $N=2$ Coulomb branch,
we find that the K\"ahler metric is given by
\beq
K_{\alpha\bar\beta}=\int_{\Sigma}U^{-1}
\omega_{\alpha}\wedge\overline{\omega}_{\bar \beta}\,,
\label{met5}
\eeq
where $\omega_{\alpha}$ are the holomorphic differentials
$\omega_{\alpha}=\w^{N_c-\alpha}\dd\w/(y-C_{N_c}(\w))$.
This is indeed different from the one of special geometry.
However, the prefactor $U^{-1}$
in the integrand is nowhere vanishing nor divergent
and is bounded from below and above as
$(1+N_fR\emst^2/2|m|)^{-1}\leq U^{-1}\leq 1$,
and hence the moduli space has the same type of singularity 
as in the $N=2$ theory. In particular, the same type
of massless particles appear as in the $N=2$ theory.
This is what is expected in field theory.
Therefore, the brane captures the correct qualitative feature of
the field theory of interest.

Let us take a closer look at this metric. The coordinates $y$ and $x$
are related to $x^6$ and $x^{10}$ by the formulae
(\ref{defY}) and (\ref{defX}). As in section~4.4, the metric depends
on the overall constants in these formulae
which is a parameter with no counterpart in field theory.
This already shows a discrepancy between the brane and field theory
results. We will only examine the metric for one particular value
that puts the D6 branes
exactly in the middle between the two NS 5-branes
in the $x^6$ direction.
Since all the D6 branes are at $x^6=0$, the requirement is
$|x|=|y|$ at $x^6=0$. This yields at $v=0$ that
$y=m^{N_f/2}\Lambda_{N=1}^{N_c-N_f/2}t$,
$x=m^{N_f/2}\Lambda_{N=1}^{N_c-N_f/2}t^{-1}$
where $|t|=1$ at $x^6=0$.
By introducing the rescaled variable $\tilx=x^6/|\elst^2m|$, 
$t$ is given by
\beq
t\,=\,\e^{-(\tilm\,\tilx+ix^{10})}
(\sqrt{1+\tilx^2}
-\tilx)^{N_f/2}\,,
\eeq
where
\beq
\tilm={|m|\over R\emst^2}\,.
\label{deftilm}
\eeq
In terms of the new variables, the curve $\Sigma$ is given by
\beq
t+t^{-1}=2C_{N_c}(\w,u_{\alpha})/(m^{N_f/2}\Lambda_{N=1}^{N_c-N_f/2})
\,,
\eeq
and the function $U$ is expressed as
\beq
U=1+{N_f\over 2\tilm}(1+\tilx^2)^{-{1\over 2}}\,.
\eeq
Then, we see that
the K\"ahler metric (\ref{met5}) depends only on
$m^{N_f}\Lambda_{N=1}^{2N_c-N_f}$,
$u_{\alpha}$, and the combination
$\tilm$ (\ref{deftilm}).

In the limit $m\to\infty$, the K\"ahler metric converges to
that of the $N=2$ Coulomb branch as is also the case for
the gauge theory of interest.
However, the precise value of $m$ above which the metric
becomes close to the one of the $N=2$ theory is different
from the one in the gauge theory.
In the gauge theory, provided $m\gg \Lambda_{N=2}$,
the metric is close to the $N=2$ metric in the region in which
$\langle\Phi\rangle \ll m$.
In the worldvolume theory of the brane, however, the metric is
close to the $N=2$ metric if $\tilm\gg 1$, namely,
\beq
m\gg R\emst^2\,,
\label{condi}
\eeq
irrespective of the values of $u_{\alpha}$.

The structure of
deviation from the $N=2$ metric looks also different.
In general
it is not easy to carry out the integral (\ref{met5}).
However, for large enough $m$ (\ref{condi}),
in the region of the moduli space where
\beq
\Lambda_{N=2}\,\ll\langle\Phi\rangle\ll\,\Lambda_{N=2}\,
\exp\left({|m|\over N_c R\emst^2}\right)\,,
\label{cond1}
\eeq
$\omega_{\alpha}\wedge\overline{\omega}_{\bar\beta}$ is
large only for $\tilx\ll 1$.
Then, we can approximate $U$ by its value at $\tilx=0$ and
the K\"ahler metric (\ref{met5}) differs from
the $N=2$ metric simply by a constant multiplication:
\beq
K_{\alpha\bar\beta}=\left(1+{N_f R\emst^2\over 2|m|}\right)^{-1}
\cdot
{g_H^2\over g_L^2}\cdot K^{N=2}_{\alpha\bar\beta}\,.
\label{met5a}
\eeq
On the gauge theory side, the deviation
is suppressed as powers of $\langle\Phi\rangle/m$
in the region where $\Lambda_{N=2}\ll\langle\Phi\rangle\ll m$.
In particular, it is not so simple as (\ref{met5a}).

Notice that both in section~4 and here the scale
$R\emst^2$ appears. In particular, it is a natural
unit in the brane construction to measure the mass
of the quark in, see (\ref{deftilm}). We briefly indicate
the origin of this scale. In the fivebrane of $M$ theory,
part of the fivebrane corresponds to the D4 brane of
Type IIA, and part corresponds to the NS5 brane of Type
IIA. The transition region between the two parts is
characterized by $\partial(\ell_{st}^2 v)/\partial x^6\sim 1$.
There are two such transition regions, separated by a 
distance $\Delta x^6$, which for pure gauge theory can
be estimated to be $\Delta x^6=-2 N_c R \log(\Lambda_{N=2}/N_c 
R\emst^2)$.
This distance $\Delta x^6$ is in the Type IIA picture to
be identified with $L_{brane}$, the distance between the
two NS branes. The coupling in the Type IIA picture was
given by $1/g^2 = L_{brane}/R$, and using the above results for
$\Delta x^6$ we see 
that $\exp(-1/g^2)=(\Lambda_{N=2}/N_c R\emst^2)^{2N_c}$.
Thus, $N_c R\emst^2$ is the scale at which the bare coupling
$1/g^2$ is defined.

\section{$N=2$ Higgs Branch}

In sections 4 and 5, we saw that the fivebrane theory on  
${\bf R}^4 \times \Sigma$ and the four-dimensional gauge
theory give quantitatively different results
for the K\"ahler metric of $N=1$ theories
and higher derivative terms of $N=2$ theory.
The possible sources of
such differences are the Kaluza-Klein modes on the non-compact surface 
$\Sigma$, which do not have counterparts in the four-dimensional gauge
theory, and the issue of the decoupling of the bulk physics.
The effect of such extra modes becomes small
in the region of the parameter
space where there is a large correction to
the eleven-dimensional supergravity approximation.
However,
the computation in the eleven-dimensional supergravity approximation
gave a correct answer for
the $N=2$ Coulomb branch metric and also for
the effective gauge coupling of the $N=2$ and $N=1$ theories
in the Coulomb phase.
In all these cases, directly or indirectly via supersymmetry,
the computation involved
the chiral two form of the fivebrane theory which has the
Lagrangian given by the second term of (\ref{fiveact}).
This suggests that this part of the fivebrane theory
is remarkably rigid as we move around in parameter space.
In this section, we perform a further test of this
observation by studying
the Higgs branches of the $N=2$ theory. The metric on these
Higgs branches is also related to the chiral two form,
and indeed in several cases we can recover the exact field
theory metric using a fivebrane calculation.

\subsection{Field Theory}

We will consider four dimensional $N=2$ supersymmetric gauge theory
with gauge 
group $SU(N_c)$ and $N_f$ hypermultiplets in the fundamental
representation (quarks). The Higgs branch (baryonic branch) 
of the theory has complex dimension
$2N_cN_f - 2(N_c^2-1)$.
There are also mixed branches (non baryonic branches) 
labeled by an integer $r$ \cite{aps} with complex dimension
$2r(N_f-r)$.
The metrics on the Higgs branches of the $N=2$ theory 
are hyper-\Kahler. The classical metric is not corrected
by quantum effects.
Therefore the exact Higgs branch metric is obtained by  
the {hyper-\Kahler} quotient construction.

In order to compare to the metric that we will compute using the fivebrane
of M theory
we have to express the metric in the coordinates of the brane construction.
In order to do that we will use the following strategy. 
 First we will consider the dimensional reduction
of the $N=2$ theory to an $N=4$ theory in three dimensions. The metric 
on the Higgs branch is not modified under the reduction. 
In three dimensions there is a mirror symmetry \cite{is} relating 
two different $N=4$ theories 
in the infrared, namely in the limit 
$1/e^2 \rightarrow 0$ where $e$ is the three dimensional gauge
coupling. Under
this mirror symmetry the Higgs and Coulomb branches of the mirror 
theories are exchanged. It turns out that the Higgs branch metric
expressed in terms of the Coulomb branch metric of the mirror in three
dimensions is naturally related to the one computed using the
fivebrane.  While the Higgs branch is not corrected quantum
mechanically the Coulomb branch is.
The metric on the Coulomb branch receives both loop and instanton corrections.
In the next subsection, we will see that these have natural
interpretations from the fivebrane point of view. 

We will start by considering the gauge group $SU(N_c)$ with $N_f=N_c$ 
hypermultiplets in the fundamental representation. 
The baryonic branch of this theory has
complex dimension two.
The effective field theory at the baryonic branch root (the point
of intersection with the
Coulomb branch)
is a $U(1)^{N_c-1}$ gauge 
theory with $2N_c-N_f$ massless electrons with charges $(-1,0,...,0)$,
$(1,-1,...,0),...,(0,...,1,-1)$, $(0,...,0,1)$ \cite{aps}.

The mirror theory is a $U(1)$ gauge theory
with $N_c$ hypermultiplets (electrons) \cite{is}.
The exact Coulomb branch metric of this theory is determined
at one loop. Higher loop
corrections are absent \cite{sw},
 while instanton corrections are absent since the gauge group is abelian.
 The exact metric on the Coulomb branch of the mirror theory takes in the
infrared limit the form 
 \beq
 ds^2 = V(\vw) d\vw^2 +  V(\vw)^{-1}(d\sigma + \vec{A} d\vw)^2
 \comma
 \label{1}
 \eeq
where
\beq
 V(\vw) = \frac{N_c}{|w|}~,~~~
 \vec{\nabla} V =  \vec{\nabla} \times \vec{A} 
 \stop
 \label{2}
 \eeq
 The computation leading to (\ref{2}) is one-loop with 
gauge fields on the external 
 legs and hypermultiplets running in the loop.
 This is the metric of an ALE space with $A_{N_c-1}$ singularity.

Consider next the gauge group $SU(N_c)$ with $N_f=N_c+1$ 
hypermultiplets in the fundamental representation.
The baryonic branch of this theory has
complex dimension $2N_c+2$.
The effective field theory at
the baryonic branch root is a $U(1)^{N_c-1}$ gauge 
theory with $2N_c$ electrons charged as \cite{aps}
\be
\begin{array}{cccccl}
&U(1)_1&U(1)_2&...&U(1)_{N_c-1}&\\
q_1&1&1&...&1&\\
q_2&1&1&...&1&\\
\vdots\\
q_{N_c+1}&1&1&...&1&\\
q_{N_c+2}&-1&0&...&0&\\
q_{N_c+3}&0&-1&...&0&\\
\vdots&\vdots&\vdots&\vdots&\vdots&\\
q_{2N_c}&0&0&...&-1&\\
\end{array}
\label{list1}
\ee
The mirror theory is a $U(1)^{N_f-N_c}=U(1)^{N_c+1}$
gauge theory with $2N_c$ electrons and charges
(after a change of basis) as follows \cite{dhooy}
\be
\begin{array}{ccccccl}
&U(1)_1&U(1)_2&U(1)_3&...&U(1)_{N_c+1}&\\
q_1&1&0&0&...&0&\\
q_2&1&0&0&...&0&\\
\vdots&\\
q_{N_c-1}&1&0&0&...&0&\\
q_{N_c}&1&-1&0&...&0&\\
q_{N_c+1}&0&1&-1&...&0&\\
\vdots\\
q_{2N_c-1}&0&0&0&...&-1&\\
q_{2N_c}&0&0&0&....&1&\\
\end{array}
\label{list2}
\ee

As in the case $N_f=N_c$, 
the exact coulomb branch metric of this theory is determined at one loop.
The metric takes the form \cite{dhooy}
\beq
 ds^2 = g_{ij}d\vw_id\vw_j +  (g^{-1})_{ij}
 (d\sigma_i + \vec{A}_{ik} d\vw_k)(d\sigma_j+ \vec{A}_{jl} d\vw_l)
 \comma
 \label{33}
 \eeq
 where 
\beqar
 \vec{\nabla}_kg_{ij} &=& \vec{\nabla}_ig_{kj} \non\\
 \frac{\pa}{\pa w_i^m} A_{jk}^n - \frac{\pa}{\pa w_j^n} A_{ik}^m &=&
 \varepsilon_{mnp} \frac{\pa}{\pa w_i^p}g_{jk}
 \stop
 \label{44}
 \eeqar 
 $g_{ij}$ has been computed in \cite{dhooy} (see eq. 4.10 there). For instance 
 the diagonal components $g_{ii}$  take the form
 \beqar
 g_{11} &=& \frac{N_c-1}{|\vw_1|} + \frac{1}{|\vw_1- \vw_2|} \non\\
 g_{ii} &=& \frac{1}{|\vw_i-\vw_{i-1}|} + \frac{1}{|\vw_{i+1}- \vw_i|}
~~~~~~~i =2,\ldots,N_c
\label{eqqq}\\
 g_{N_c+1,N_c+1}&=&{1\over |\vw_{N_c}-\vw_{N_c+1}|}
+{1\over |\vw_{N_c}|} 
\nonumber
 \eeqar
As in the previous case, there are no higher loop or instanton
 corrections to the metric. 

The same analysis as above can be repeated for the $r=1$
non-baryonic branch for general $N_f$ and $N_c$.
In this case
the effective gauge theory at the root is
abelian and the exact
metric can be determined by one loop computation of the Coulomb branch
of the mirror theory.
(The theory at the root is actually the mirror of the theory
at the baryonic branch root of the $N_f=N_c$ theory considered above.)

 For other cases, the story is much more involved.
Consider the $r>1$ non baryonic branch.
The effective theory at a generic point of the root is
$U(r)$ with $N_f$ 
hypermultiplets in the fundamental representation
(plus a free Maxwell theory).
The mirror gauge theory has gauge group
$U(1)\times U(2)\times\cdots\times U(r-1)\times
U(r)^{N_f-2r+1}\times U(r-1)\times \cdots \times U(1)$,
with hypermultiplets
in the bi-fundamentals of each two adjacent groups and
a hypermultiplet in the fundamental representation
of each of the first and the last $U(r)$ \cite{hw,hov}.

Since the theory is non abelian, 
the one-loop metric receives contributions both from hypermultiplets
and vector multiplets running in the loop.
Let us label the unitary groups by $i,j,...=1,\ldots,N_f-1$
and by $a_i,b_i,...=1,\ldots,k_i$
the color indices of the $i$-th group $U(k_i)$.
The Coulomb branch of the mirror theory is parametrized by
$\vec{w}_{i a_i}$ and $\sigma_{i a_i}$.
While the hypermultiplet in the $(i\bar{j})$ bi-fundamental  
contributes $1/|\vw_{i a_i}-\vw_{j b_j}|$
to the $g_{i a_i, i a_i}$ component of
the metric, a vector multiplet 
contributes $-2/|\vw_{i a_i}-\vw_{i b_i}|$ \cite{dhoo}, which
becomes large negative
in the limit $\vw_{i a_i}\to\vw_{i b_i}$. However, there are also
instanton corrections relevant in this region coming from
the 't Hooft-Polyakov monopole, which are typically of order
\beq
\e^{-{|\vw_{i a_i}-\vw_{i b_i}|\over e^2}}\,.
\label{FTinst}
\eeq
These render the full metric positive definite.

\subsection{Fivebrane Back Reaction}
\newcommand{\volp}{{\rm vol}({\bf P}^1)}

In the Higgs branch, the fivebrane worldvolume has several disjoint
components.
In the IIA pictures, they are segments of D$4$ branes
stretched between D$6$ branes and their positions parametrize the
Higgs branches of the $N=2$ theory. In the M theory picture, these
are fivebranes wrapping the ${\bf P}^1$ cycles
on the Taub-NUT space \cite{witten1,hoo}.
The Higgs branch coordinates are
then the locations (labeled by $\vec{w}$)
of the ${\bf P}^1$ parts of the fivebrane transverse
to the Taub-NUT geometry and vevs (labeled by $\sigma$) of the
self-dual two-form $B$ on the ${\bf P}^1$'s. They combine to make
hyper-K\"ahler coordinates appropriate for the $N=2$ Higgs branch. 

Let us pay attention to one of such ${\bf P}^1$ parts of the fivebrane
and define
\be
  (x^7,x^8,x^9) = \ell_{11}^3 \vw,~~
  B_{z\bar{z}} = \sigma \frac{\omega_{z\bar{z}}}{\volp} .
\label{scaling}
\ee
To do the Kaluza-Klein reduction on the fivebrane worldvolume, we
regard $\vw$ and $\sigma$ as functions of $x^{0,...,3}$ and set 
$\omega_{z\bar{z}}$ to be the volume form on ${\bf P}^1$. We have
chosen the normalization of $\sigma$ so that its periodicity is
$1$.  If we use the flat metric for the spacetime, the fivebrane effective
action for $\vw$ and $\sigma$  would simply be
\be
  \int d^4 x  
\Big( \volp \partial^m \vw \partial_m \vw + \frac{1}{\volp}
 \partial^m \sigma
  \partial_m \sigma \Big),
\ee
and does not correctly reproduce the Higgs branch metric expected 
from the field theory analysis. In the low energy approximation of 
M theory, the only way that other components of the fivebrane 
affect the kinetic term for
the ${\bf P}^1$ is through the graviton exchange.

As a warm-up exercise, let us compute the back reaction effect when
the size of ${\bf P}^1$ is larger than the distance between the
branes.
In this limit, the fivebrane configuration is almost flat. If a
fivebrane is stretched in the $x^{0,...,3}, x^{4,5}$ directions, the metric
induced by it is 
\bea
 ds^2 &=& g_{\mu\nu} dx^\mu dx^\nu \nonumber \\
 &=& \left(1 + \frac{\ell_{11}^3}{r^3}\right)^{-1/3} \sum_{i=0}^5 (dx^i)^2 +
      \left(1+ \frac{\ell_{11}^3}{r^3}\right)^{2/3} \sum_{i=6}^{10} (dx^i)^2,
\label{backreaction}
\eea
where
\be 
  r = \sqrt{\sum_{i=6}^{10} (x^i)^2}.
\ee
and the horizon of the fivebrane is located at $r=0$. We then consider
another fivebrane located at $(x^7,x^8,x^9) =
\ell_{11}^3 \vw$ and stretched in the
$x^{0,...,3},x^{6,10}$ directions. To derive the kinetic term for
$\vw$, we take $\vw$ to be a function of $(x^0,...,x^3)$ and
substitute it into the fivebrane action:
\be
  S_{\vw}=  
 \int dx^0 \cdots dx^3
 \int dx^6 dx^{10} \sqrt{{\rm det}G} \sum_{m,n=0}^3
   \sum_{a,b=7,8,9} G^{mn} g_{ab} \partial_m w^a \partial_n
 w^b,
\ee
where  
$g_{ab}$ is the spacetime metric given by (\ref{backreaction}) and
$G_{mn}$ is the induced metric on the fivebrane. This integral can be
evaluated explicitly and one finds,
\bea
S_{\vw} &=& \int d^4 x \int dx^6 dx^{10}
  \left(1 + \frac{\ell_{11}^3}{(\sum_{i=6}^{10} (x^i)^2)^{3/2}} \right)
   \partial^m \vw \cdot \partial_m \vw \nonumber \\
&=& \int d^4x \left( \volp  + \frac{\pi/2}{|w|} \right)
    \partial^m \vw \cdot \partial_m \vw 
\label{waction}
\eea   
Note that the $\ell_{11}$ dependence disappears from
the final expression. 

The kinetic term for $\sigma$ comes from 
\be
    S_{B} = \int d^6x |dB - C^{(3)}|^2
\label{sigmaaction}
\ee
of the fivebrane action, so we have to evaluate the 3-form potential 
$C^{(3)}$ induced by the fivebrane at $r=0$, i.e.
\be
  d C^{(3)} = \frac{1}{r^5} \sum_{ijklm=6}^{10} x^i dx^j \wedge
dx^k \wedge dx^l \wedge dx^m
\ee 
Up to a suitable gauge choice, we can solve this as
\bea
   C^{(3)}_{a=7,8,9; ij=6,10} &=& A_a \frac{y^3}{r^5} \epsilon_{ij} 
\nonumber \\ 
 C^{(3)}_{a,b=7,8,9; i=6,10} & = & x_{[a,} A_{b]} 
 \frac{1}{r^5} 
 \epsilon_{ij} x^j \nonumber \\
 C^{(3)}_{789} &=&0 ,
\label{aux7}
\eea
where $A_i$ is the magnetic monopole vector potential in the
$x^{7,8,9}$ space,
\be
  A_i dx^i = ({\rm cos}\theta \pm 1) d \phi,
\ee
and $y = \sqrt{\sum_{a=7,8,9} (x^a)^2}$. 
On ${\bf P}^1$ where the fivebrane is wrapped, 
$A_a$ is constant and $y^3/r^5$ is proportional to the volume form
$\omega_{z\bar{z}}$ up to a cohomologically
trivial form. Thus, with a suitable shift of the 2-form $B$ in the
definition of $\sigma$ in (\ref{scaling}), we find
\bea
 (dB - C^{(3)})_{|{\bf R}^4 \times {\bf P}^1} 
  &=& 
  (d \sigma -
  A_i d x^i)_{|{\bf R}^4} \wedge
\frac{\omega_{z\bar{z}}}{\volp} .
\label{aux9}
\eea
Notice that when we consider $dB-C^{(3)}$ on ${\bf R}^4 \times
{\bf P}^1$, there are also terms coming from the second line in 
(\ref{aux7}), but these give rise to higher derivative terms 
in the action and have been dropped in (\ref{aux9}). The three-form
in (\ref{aux9}) is not yet self-dual. To make it self-dual, we add
to $B$ the two-form
\be
B_0 = 
 \left(1 +
\frac{1}{r^3} \right)^{-1}
   \frac{1}{\volp}
 x_n(\partial_m \sigma -
  A_i \partial_m x^i)
(dx^n \wedge dx^m + \epsilon^{nm}{}_{pq} dx^p \wedge dx^q)
\ee
where $m,n,p,q \in \{0,1,2,3\}$. Again dropping higher derivative
terms, $dB_0$ contains two pieces. One of them is the dual of (\ref{aux9}),
the other is a self-dual three-form. Thus, $d(B+B_0)- C^{(3)}$ is
a self-dual three-form. In addition, the gauge fixing conditions
$d^\ast B=0$ and $d^\ast C^{(3)}=0$ are satisfied. This is precisely the
situation as described in section~2, showing that we should do our
computations with the three-form $d(B+B_0)-C^{(3)}$. Luckily,
$\int |d(B+B_0)-C^{(3)}|^2 = 2\int  |dB-C^{(3)}|^2$, so we do not make an
error is we work with just the three-form (\ref{aux9}).

The action is then evaluated as
\bea
 S_{\sigma} &=& \int d^4x \int \frac{ dx^6 dx^{10}}{(\volp)^2} 
\left(1 +
\frac{\ell_{11}^3}{(\sum_{i=6}^{10} (x^i)^2)^{3/2}} \right)^{-1}
 \left( \partial_m \sigma -
A_i(w) \partial_m w^i \right)^2 \noindent \\
&=& \int d^4x \left( \frac{1}{\volp} - \frac{\pi/2}{(\volp)^2 |w|} 
+ \cdots
\right) \left( \partial_m \sigma -
A_i(w) \partial_m w^i \right)^2.
\eea 
Combining this with (\ref{waction}), we find that the metric on the
moduli space of the fivebrane wrapping on ${\bf P}^1$ is given by
\be
 ds^2 = V(w) d\vw^2 + V(w)^{-1} (d \sigma - A_i dw^i)^2
\label{standard}
\ee
with
\be
  V(w) = \volp + \frac{\pi/2}{|w|} 
\ee
and
\be
  \nabla V  =  \nabla \times A .
\label{hyperkahlercond}
\ee

Although we have obtained (\ref{standard}) by computing each term
explicitly, we could have gotten it faster by using the following argument.
It is known that (\ref{standard}) with $V$ and $A$ obeying
(\ref{hyperkahlercond}) is the most general hyper-K\"ahler metric for
$(\sigma, \vw)$ with the rotational symmetry for $\vw$ and the
translational symmetry for $\sigma$. The condition
(\ref{hyperkahlercond}) requires $\Delta V = 0$ almost
everywhere except for singularities in $V$. In the present case, a
 singularity comes only from $\vw =0$. The vector potential $A$
is then uniquely determined by the intersection of the two components
of the fivebranes, one stretching in the $x^{4,5}$ directions and the
other stretching in the $x^{6,10}$ directions. This in turn fixes
$V$ up to an additive constant. 
 
This argument is also applicable when the size of ${\bf P}^1$ is comparable
to $x^{7,8,9}$ or smaller. In this case, we can no longer ignore the effect of
the background Taub-NUT geometry. The metric induced by the fivebrane
wrapping on cycles on the Taub-NUT space is not known
explicitly. Nevertheless it is still true that the resulting moduli
space metric in the supergravity limit should be of the form
(\ref{standard}) with $V$ and $A$ obeying
(\ref{hyperkahlercond}). This is because the rotational symmetry in
the $x^{7,8,9}$ plane is unaffected by the Taub-NUT geometry and
the translational invariance in $\sigma$ is unbroken in the
supergravity limit. It is then sufficient to determine $A$,
which can be found as follows. 

Suppose the
fivebrane is wrapped on a surface ${\bf R}^4 \times \Sigma$ at the
origin of the $x^{7,8,9}$ plane, where
$\Sigma$ is a cycle on the Taub-NUT space. Let $C^{(3)}$ be the 3-form
potential induced by such a fivebrane. Then for any other cycle
$\Sigma'$ on the Taub-NUT space,
\be
   \int_{S^2 \times \Sigma'} dC^{(3)} = \# (\Sigma \cap \Sigma')
\ee
where $S^2$ surround the origin in the $x^{7,8,9}$ plane and
$\# (\Sigma \cap \Sigma')$ is the intersection number of the two
surfaces. This means
\be
   C^{(3)} = \frac{\# (\Sigma \cap \Sigma')}{{\rm vol}(\Sigma')}
   A_i dx^i \wedge
    \omega_{z\bar{z}} + \cdots
\ee 
where $A_i$ is the vector potential for the monopole at the origin of
the $x^{7,8,9}$ plane and $\omega_{z\bar{z}}$ is the volume form on 
$\Sigma'$. This then determines $A$ for the moduli space
metric. 

It is straightforward to generalize this construction when there are
several disjoint ${\bf P}^1$'s. In this case, the metric takes the
form (cf. (\ref{33}) and (\ref{44}))
\beq
 ds^2 = g_{ij}d\vw_id\vw_j +  (g^{-1})_{ij}
 (d\sigma_i + \vec{A}_{ik} d\vw_k)(d\sigma_j+ \vec{A}_{jl} d\vw_l)
 \comma
 \eeq
 where 
\beqar
 \vec{\nabla}_kg_{ij} &=& \vec{\nabla}_ig_{kj} \non\\
 \frac{\pa}{\pa w_i^a} A_{jk}^b - \frac{\pa}{\pa w_j^b} A_{ik}^a &=&
 \varepsilon_{abc} \frac{\pa}{\pa w_i^c}g_{jk}
 \stop
 \label{4}
 \eeqar 
Here $\vw_i$ specifies the location of the $i$-th ${\bf P}^1$ and
$\sigma_i$ is the vev of the self-dual $B$-field on it. By
generalizing the above argument, one can show that 
$A_{ij}$ is uniquely determined by the intersection of the $i$-th
and the $j$-th ${\bf P}^1$. In the Taub-NUT space of $A_N$ type, the
intersection number of neighboring cycles on the Taub-NUT space is
$+1$ while the self-intersection number of each cycle is $-2$. 
Thus, with the identification
$1/e^2 = \volp$, one can easily see that
the metric computed in this way agrees with the field
theory result in the previous subsection, up 
to the instanton corrections.

To obtain the four-dimensional gauge theory from the fivebrane, we
should be able to neglect the Kaluza-Klein modes. Since the typical
energy scale of the gauge theory is $|w|$, we need
\be
    |w| \ll \frac{1}{\volp}.
\label{ineqq}
\ee
This is the same as taking the infrared limit $1/e^2 \rightarrow 0$ in
the field theory analysis in the previous subsection.
This is also the limit where the $N_f$ D6-branes, separated in the $x^6$
direction, become close to one another and the $SU(N_f)$ symmetry
is restored.
This is indeed the limit where the worldvolume theory becomes
close to the gauge theory since
the latter has the $SU(N_f)$ flavor symmetry. 
In the spacetime coordinates related to $\vw$ by (\ref{scaling}),
(\ref{ineqq}) means
\be
   |x^{7,8,9}| \volp \ll \ell_{11}^3.
\label{stronggravity}
\ee
One can expect that quantum corrections to the supergravity
description become strong in such a situation.\footnote{It was shown in
\cite{ght} that the fivebrane metric (\ref{backreaction}) is smooth
across the horizon at $r=0$. This however is not the case when the
fivebrane is wrapped on a compact surface such as ${\bf P}^1$
and the curvature is
expected to diverge near the horizon. We thank G. Horowitz and
J. Maldacena for discussion on this issue.}
To see this more
explicitly, let us consider the case when there are two components
of the fivebrane wrapping on two copies of a ${\bf P}^1$ of the
Taub-NUT space
separated in the $x^{7,8,9}$ directions. As a membrane can end on the
fivebrane, a membrane stretched between the two ${\bf P}^1$ may cause
instanton effects. The membrane action measured with respect to the
metric (\ref{backreaction}) would be proportional to
$\volp |\Delta x^{7,8,9}|/\ell_{11}^3=\volp |\Delta\vec{w}|$
and thus the instanton 
effects are typically
\beq
\e^{-\volp|\Delta\vec{w}|}
\label{openM2}
\eeq
which become large in the limit (\ref{ineqq}) or (\ref{stronggravity}). 
The instanton would break the translational
invariance in $\sigma$, and thus the statement in the above paragraph
does not hold in such a case. By following a chain of duality
arguments, one can show that these membrane instantons are in
one-to-one correspondence to the field theory instantons discussed in
the previous subsection
and have the same effect on the metric
(compare (\ref{openM2}) with (\ref{FTinst})).

So far we have been considering the variation of the $x^{7,8,9}$
position and the chiral two form
of the finite ${\bf P}^1$ components. We have shown that
the brane computation correctly reproduces
the field theory result for non-baryonic branches (up to the membrane
instanton correction). However, for the baryonic branches,
there is one quaternionic modulus in addition to such
${\bf P}^1$ motions.
As discussed in \cite{hoo},
this is identified with the one modulus associated with
the charged massless particles which appear when the infinite
component factorizes into two components.
It is natural to guess that this is related to the relative separation
of the two components.
However, since both components are of infinite volume,
a na\"\i ve computation shows that
it costs an infinite energy to separate them
and also it is not clear whether there is a zero mode for
the chiral two form.
Nevertheless, if we assume that there is one zero mode for the
chiral two form and there is a way to make the kinetic energy finite,
the back reaction method again gives us the correct field theory
result. Note that the two infinite components
intersect at $2N_c-N_f$ points \cite{hoo}.
In the case $N_f=N_c$, this yields
the equation $V=N_c/|\vw|$ in (\ref{2}).
In the case $N_f=N_c+1$, $\vw_1$ is interpreted as
the location of
one of the infinite components and the other is fixed at $\vw=0$.
Then, the first and the last equation in (\ref{eqqq}) follow
from the fact that
the two infinite components intersect at $N_c-1$ points, and that
one of them intersects 
the first ${\bf P}^1$ 
at one point whereas
the other intersects the last ${\bf P}^1$ at one point.

\medskip
\noindent
{\it Remarks}.\\
(1) The 
Higgs branch metric has also been discussed in the framework of
geometric engineering in \cite{GMV}
where the Higgs branch of an $U(1)$ theory was considered.\\
(2) We also note that the above computation can be extended to
$Sp(N_c)$ gauge theories
and compactified 6d tensionless string theory
by using the configuration constructed in \cite{hov}.
In these cases, open membrane instantons always contribute.

\section{$N=1$ Higgs Branch}

We consider four-dimensional $N=1$ supersymmetric gauge theory
in the phase where the gauge group is broken by the
Higgs mechanism.
We consider two kinds of theories.
One is obtained from $N=2$ SQCD with
$SU(N_c)$ group and $N_f$ flavors
by giving mass $\mu$ to the adjoint chiral multiplet.
The other is $N=1$ supersymmetric QCD. The fivebrane
configuration in $M$ theory corresponding to
these theories has been obtained in \cite{hoo}.
We consider the simplest example $N_c=N_f=2$ in some detail.

In both cases, we use the back-reaction method to compute the metric
on the moduli space as in the computation of the metric of the
$N=2$ Higgs branch. 
In the case of $N_f=N_c=2$,
there are no complications associated with open membrane
instantons.

\subsection{$N=2$ Broken to $N=1$ by Adjoint Mass}

\subsection*{\sl Field Theory}

$N=2$ gauge theory with $SU(2)$ gauge group and $N_f=2$ fundamental
hypermultiplets has been studied in detail in \cite{sw1}.
The moduli space consists of a complex one-dimensional Coulomb branch
and two Higgs branches emanating from different points on the Coulomb
branch. One of the Higgs branches is of baryonic type and the other is
of non-baryonic type. These are both isomorphic to
an Eguchi-Hanson space with an $A_1$ singularity at the point of
intersection with the Coulomb branch.

If we give a bare mass
$\mu$ to the adjoint chiral multiplet, $N=2$ supersymmetry
is broken to $N=1$ and all but the two Higgs branches are lifted.
The Higgs branch is K\"ahler but
no longer a hyper-K\"ahler manifold,
and it is difficult to compute the K\"ahler metric because it
received both loop and instanton corrections.
However,
it is easy to see how the complex structure of these Higgs branches
is deformed as a function of $\mu$.
Here we only consider the non-baryonic branch.
The non-baryonic branch is parametrized by the meson vev
$M=\tilQ Q$ which is a $2\times 2$ complex matrix.
As analyzed in \cite{hoo},
a general meson vev can be made by a flavor rotation
into a diagonal matrix
with eigenvalues $(i\mu\Lambda, -i\mu\Lambda)$ 
where $\Lambda$ is the dynamical scale of the $N=2$ theory
(therefore the moduli space is set-theoretically
the homogeneous space $SL(2,\C)/\C^*$).
Thus, at a general point of the moduli space,
the meson matrix is expressed as
\beq
M=g\left(
\begin{array}{cc}
i\mu\Lambda&0\\
0&-i\mu\Lambda
\end{array}
\right)g^{-1}=i\mu\Lambda
\left(
\begin{array}{cc}
ad+bc&2ab\\
2cd&-ad-bc
\end{array}
\right)
\,,\qquad ad-bc=1\,.
\eeq
In terms of the variables $M_{11}=W$, $M_{12}=X$, $M_{21}=Y$,
the moduli space is described as a complex manifold by
\beq
XY=(W-i\mu\Lambda)(W+i\mu\Lambda)\,.
\label{split}
\eeq
Namely, the $A_1$ singularity of the $\mu=0$ Higgs branch
has split into two $A_0$ ``singularities''
and is smoothed out. Although this complex manifold
admits a hyper-K\"ahler metric, we do not expect
the metric of the moduli space to be hyper-K\"ahler.

\subsection*{\sl Fivebrane Back Reaction}

In \cite{hoo}, we have seen that
the fivebrane correctly reproduces the relation of the
flavor invariants of supersymmetric QCD with and without heavy
adjoint.
There we also proposed that the flavor
rotation is related to the position of the $\CP^1$ components
by counting the number of degrees of freedom.
However, the precise map between the position of the
$\CP^1$ components and the points in the
flavor orbit has not been given.
Here we compute the metric of the non-baryonic branch by the
back-reaction method and read off its complex structure
to compare with (\ref{split}).

\begin{figure}[htb]
\begin{center}
\epsfxsize=4in\leavevmode\epsfbox{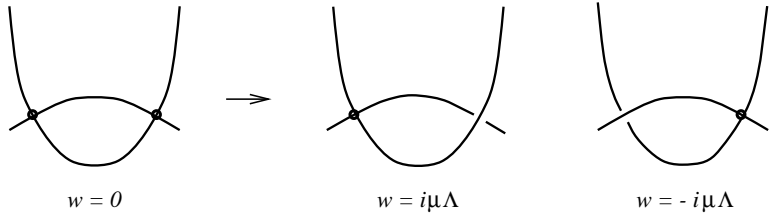}
\end{center}
\caption{Splitting $A_1$ Singularity to $A_0\times A_0$
by Adjoint Mass}
\label{splitfig}
\end{figure}

As shown in \cite{witten1,hoo}, the fivebrane at this non-baryonic
branch consists of two components --- an infinite component $C$ and
a finite component ${\bf P}^1$. The position $\vec{w}$
of the ${\bf P}^1$
component in the $789$ directions and the integral $\sigma$
of the chiral
two form on ${\bf P}^1$ correspond to
the real four dimensions of this branch.
 For $\mu=0$, the ${\bf P}^1$ component intersects the infinite
component $C$ at two points at the origin $\vec{w}=0$ of the
$789$ direction \cite{witten1}.
 For $\mu\ne 0$,
the infinite component $C$ is rotated in the 45-89 plane,
and
the ${\bf P}^1$ component intersects $C$
at one point at one value of $\vec{w}$, $(w^7,w=w^8+iw^9)=(0,i\mu\Lambda)$,
and at one point at another value $(w^7,w)=(0,-i\mu\Lambda)$
\cite{hoo} (see Figure \ref{splitfig}).

The motion of the ${\bf P}^1$ component is affected by the
infinite component $C$ and the effect on the metric is localized
at these intersection points when all characteristic scales
are much larger than the eleven-dimensional Planck scale. 
As shown in the previous section,
the metric for $\mu=0$ is given by
\beq
\dd s^2= U(\dd \vec{w})^2+U^{-1}(\dd\sigma +\vec{A}\cdot \dd\vec{w})^2
\label{metmu}
\eeq
where
\beqa
&&{\rm curl}\,\vec{A}={\rm grad}\, U\\
&&U={\rm vol}({\bf P}^1)+{2\over |\vec{w}|}
\eeqa
The numerator ``2'' in the expression for $U$ shows that the
${\bf P}^1$ component intersects $C$ at two points at the same
time, and indicates the $A_1$ singularity.

In the case of $\mu\ne 0$, the two intersection points split into
one intersection point at $w=i\mu\Lambda$ and one intersection
point at $w=-i\mu\Lambda$ (both at $w^7=0$).
Moreover, the infinite component $C$ is sloped in the $v$-$w$
directions as:
\beq
w\,\sim\, \pm \,{\mu\over 2}\, v
\eeq
at the intersection point $w=\pm i\mu\Lambda$.
Then, a na\"\i ve back-reaction method would show that
the metric of this branch is
given by (\ref{metmu}) in which
${\rm curl}\,\vec{A}={\rm grad}\, U$ and $U$ is given by
\beq
U={\rm vol}({\bf P}^1)
+{1\over\sqrt{c^2 |w-i\mu\Lambda|^2+(w^7)^2}}
+{1\over\sqrt{c^2 |w+i\mu\Lambda|^2+(w^7)^2}}
\eeq
where $c^2$ is a constant of order $\sim 1/(1+|R\mu/2|^2)$.
This is only an approximation since there is no reason why
the coefficients of $(\dd\vec{w})^2$ and $(\dd \sigma+\cdots)^2$
should be the same in the case where $N=2$ supersymmetry is broken to
$N=1$. However, not knowing the correct way to carry out
the computation, we examine this na\"\i ve approximation
of the metric.

There is a complex structure with respect to which this metric
is {\it hermitian}. One of the complex coordinates is $w$. Another
coordinate is given by
\beq
y=\e^{-({\rm vol}({\bf P}^1)w^7+i\sigma)}\prod_{\epsilon=\pm i}
\sqrt{\sqrt{c^2|w-\epsilon \mu\Lambda|^2+(w^7)^2}
-w^7}\,\cdot\,{\rm const}\,.
\label{defy}
\eeq
Then, the metric is expressed as
\beq
\dd s^2=U\,|\dd w|^2
+\,U^{-1}|{\dd y/y}-\delta \dd w|^2\,,
\eeq
where 
\beq
\delta={1\over 2}\sum_{\epsilon=\pm i}{1\over w-\epsilon\mu\Lambda}
\left(1+{w^7\over\sqrt{c^2|w-\epsilon \mu\Lambda|^2+(w^7)^2}
}\right)\,.
\eeq
Thus, the metric is indeed hermitian.
There is yet another complex coordinate
\beq
x=\e^{{\rm vol}({\bf P}^1)w^7+i\sigma}\prod_{\epsilon=\pm i}
\sqrt{\sqrt{c^2|w-\epsilon \mu\Lambda|^2+(w^7)^2}
+w^7}
{w-\epsilon\mu\Lambda\over |w-\epsilon\mu\Lambda|}\,
\cdot\,{\rm const}.
\label{defx}
\eeq
Then, $x$, $y$ and $w$ satisfy the relation
\beq
xy={\rm const}\cdot (w-i\mu\Lambda)(w+i\mu\Lambda)\,,
\eeq
and this describes
the complex structure of the moduli space with respect to which
the metric is hermitian.
This agrees with the field theory result (\ref{split}).
Thus, we propose that the ${\bf P}^1$ motion
is mapped to the flavor rotation by $x=X$, $y=Y$,
and $w=W$.

However, there is a serious problem.
Since the theory is $N=1$ supersymmetric, the moduli space of vacua
must be K\"ahler. Although the above metric is
hermitian, it is not K\"ahler
with respect to the above chosen complex structure.
It is not clear whether this is
because of the too na\"\i ve approximation
or because of the wrong choice of the complex structure.
We leave it as an open problem.
If we can show that the reason is the former but not the latter,
it would support the above refinement of the proposal
about the
relation of the flavor rotation and the ${\bf P}^1$ motion.

In principle, the correct
complex structure could be identified by looking
at the action of the supersymmetry on the worldvolume fields
or on the parameters $(\vec{w},\sigma)$.
Although it is not clear whether it is practical,
it is important to show that the
complex structure of the moduli space
is independent of the extra parameter $R$.
It is clear that the part of the complex structure
which is directly induced from the complex structure
of the space-time (such as $w=w^8+iw^9$)
does not depend on the extra parameter $R$,
but it is less clear whether it is true for the part
related to parameters such as $(w^7,\sigma)$.\footnote{
It is interesting to note that
the pair $(w^7,\sigma)$ looks very much like the
scalar fields one gets after dualizing the 
linear multiplet of $N=2$
theory in three dimensions.
The expression (\ref{defy})
or (\ref{defx}) is almost the same as the expression for the
superpotential (which is a holomorphic function) that
appeared in the study of three-dimensional $N=2$ gauge theory
(see (3.11) in \cite{dho}).}

\subsection{$N=1$ SQCD}

We will consider four dimensional $N=1$ supersymmetric
gauge theory with gauge group $SU(2)$
and $N_f=2$ pairs of chiral multiplets
in the fundamental and anti-fundamental representations.
The moduli space of vacua is a K\"ahler manifold
whose complex structure is corrected by a
non-perturbative effect.
The classical metric is corrected
by quantum effects both perturbatively and non-perturbatively.
As in the $N=2$ case, the comparison to the brane computation
becomes more transparent if
we express the metric by compactifying on a circle to three dimensions
and go to the mirror gauge theory.
The compactification on a circle of an $N=1$ theory in four dimensions
is a three-dimensional $N=2$ theory.
The classical metric on the Higgs branch
of the four dimensional $N=1$ theory
is identical to the classical metric of the three dimensional
$N=2$ theory.
In a certain class of three-dimensional $N=2$ theories,
there is a mirror symmetry relating two different theories 
in the infrared \cite{dhoy}, namely in the limit 
$1/e^2 \rightarrow 0$ where $e$ is the  three dimensional gauge
coupling.
Under this mirror symmetry
the Higgs and Coulomb branches of the mirror theories are exchanged.
Although both the Higgs
and Coulomb branches get corrected quantum mechanically,
the one loop correction to the Coulomb branch metric
is seen classically on the Higgs branch of the mirror.
We will show that the
metric obtained using the back reaction method 
in the previous section is close
to the one loop metric on the Coulomb branch of the mirror theory.
Instanton corrections to the Coulomb branch metric correspond to the
membranes wrapping the three cycles which we discussed i
in the previous section.

The Higgs branch of this theory has
complex dimension five.
The M theory  description  of the Higgs branch was studied in \cite{hoo}.
There are two complex moduli $(\vw,\sigma)$ associated with  
the motion of the $P^1$ in $(x^7,x^8,x^9)$
and the integral of the chiral 2-form on the $P^1$.
This corresponds to the motion
of a D4 brane broken between two D6 branes and the $A_6$ gauge field component
on the D4 brane in the Type IIA picture.
 Two 
 complex moduli $m_1,m_2$ 
 are associated with the deformation of an infinite component $C_L$ 
 of the curve in $(x^8,x^9)$. These  correspond 
 to the motion of two D4 branes  broken between a  D6 brane and the NS' brane
  in the Type IIA picture.
 One complex 
 modulus $n$ has its real part 
 associated with the relative motion in $x^7$  of the infinite components
 of the curve $C_L$ and $C_R$,
 and its imaginary part is associated to the integral of the chiral
2-form on $C_L$ and $C_R$.
 The real part of
 $n$ corresponds in the Type IIA description to a relative motion of
 the NS and NS' branes in $x^7$. 
A computation as in \cite{witten1} 
implies that an infinite energy is needed for the motions $n$ and $m_1+m_2$.
 By considering them as  moduli
 we are assuming that a more elaborate computation
will show that only  finite energies  are needed for them.

 The mirror theory is a $U(1)$ gauge theory with matter content of $N=4$ 
 gauge theory with two charged hypermultiplets and a meson $M$ that couples
 to the charged matter as  $M_i \tilde{q}_i q_i$.\footnote{This mirror
 symmetry is approximate and it only captures a tree level coupling of 
 the meson to the quarks and therefore cannot be
used to get an exact metric on the
 moduli space of vacua.}
  $(\vw,\sigma)$  are the vev's for the three real
scalars in the vector multiplet
  and the dual to the gauge field respectively.
  In the three dimensional gauge theory $n$ is a parameter rather
  than a modulus, its real part is the FI parameter.
 Following \cite{dhoy} the one loop metric
 on the Coulomb branch of the mirror theory takes the form 
 \beqar
 ds^2 = \left(\frac{1}{|\vw + \vm_1|} + \frac{1}{|\vw + \vm_2|} \right)
 d\vw^2 +  \left(\frac{1}{|\vw + \vm_1|}
+ \frac{1}{|\vw + \vm_2|} \right)^{-1}
  (d\sigma + \cdots)^2 \non\\
 + \frac{1}{|\vw +\vm_1|}d m_1 d\bar{m_1} +
\frac{1}{|\vw+\vm_2|}d m_2 d\bar{m_2}  
 \comma
 \label{met}
 \eeqar
 where $\vm_i = (0,m_i)$.
 The term $\left(\frac{1}{|\vw + \vm_1|} +
\frac{1}{|\vw + \vm_2|} \right)$
 is a consequence of the fact that
 in the mirror theory the moduli $m_1,m_2$ correspond to a meson that couples
 to the charged hypermultiplet  as $m_i \tilde{q}_i q_i$. 
 The ``$\cdots$'' part
depends on $d \vw, d m_1, d m_2$
and is determined by the duality transformation
 of the vector to a scalar in the presence of the meson, such that
 the metric (\ref{met}) is
 {\Kahler}.
 As we noted previously, the
  four dimensional  modulus $n$ is a parameter
  in three dimensions.

 The metric (\ref{met}) is  obtained via an approximate back reaction,
 where the infinite component $C_L$ is approximated by a straight brane and the
 back reaction of the infinite component $C_R$ is neglected.
 Therefore this classical supergravity approximate back reaction reproduces 
 the one loop metric of the 
 Coulomb branch of the mirror theory.
 This however only captures  classical features of the
  metric on the Higgs branch of the original theory.
 We  expect that the classical metric will be corrected by loops.
These higher loops cannot
be  seen in the appproximate back reaction that we used.
It is probable that an improved back reaction method that takes 
into account the structure
of the infinite component $C_L$ as well as $C_R$ will capture the higher loops.

The example above can be generalized in a straightforward way 
to general $N_f,N_c$. 
The modifications compared to the $N=2$ case 
are the meson moduli that add mass terms for  some 
of the quarks \cite{dhoy}. However, in this case we also 
expect instanton corrections
and we have to compute the membrane instanton
contribution.

\section{Discussion}

The worldvolume theory
on the $M$ theory fivebrane wrapped
on a non-compact Riemann surface
depends on a scale, the radius $R$ of the circle in the eleventh
direction, in addition to the parameters that can be identified with
the parameters or the moduli of the four-dimensional gauge theory.
It is the presence of this additional parameter that
simplifies the analysis.

In general, the worldvolume theory becomes close to a standard
four-dimensional gauge theory in the region where $R\ll \elel$
and some of the characteristic lengths
of the brane become small compared to $\elel$
because the degrees of freedom absent in the four-dimensional
theory becomes heavy.
On the other hand, $M$ theory is well-approximated by the
eleven-dimensional supergravity in the opposite region where
$R\gg\elel$ and all the characteristic lengths
of the brane becomes much larger than $\elel$.

However, some quantities do not depend on
the additional scale $R$ and can be computed by going to
the region of the parameter space where the eleven-dimensional
supergravity approximation is valid.
As has been shown already, examples of such quantities are
some flavor invariant combination of the vev of
gauge invariant chiral operators,
the mass of BPS particles and the tension of BPS domain walls.
It seems that all the BPS or holomorphic quantities 
fall into this class.
It is important to understand whether and why this assertion holds,
and whether any exceptions can be found.

In this paper, we have computed the low energy effective
action of the worldvolume theory in the region of the
parameter space where the supergravity approximation is valid.
In particular, we have computed K\"ahler metric of some
$N=2$ and $N=1$
theories and higher derivative terms of $N=2$ Coulomb branch.
In the case of the $N=2$ Coulomb branch metric, the effective
gauge coupling of $N=1$ and $N=2$ abelian Coulomb phase,
and the Higgs branch metric of some $N=2$ theories,
the supergravity computation has lead to a correct result of
the corresponding four-dimensional gauge theory.
In all these successful cases, the computation involves
the chiral two form on the fivebrane,
directly or indirectly via supersymmetry,
where its lagrangian in the supergravity limit is
given by the second term of (\ref{fiveact}).
This indicates that this part of the worldvolume theory
of fivebrane has a remarkable property
that the supergravity approximation is valid
in a wider region of the parameter space than na\"\i vely
expected.

This does not apply to the calculation of non-holomorphic quantities
such as the {\Kahler} potential of $N=1$ supersymmetric gauge theories
or the higher derivative terms in the effective action
of $N=2$ theories, where the computation is
disconnected from the chiral two-form piece.
We found that the results are quantitatively strikingly different
than what we expect for the four dimensional gauge theories.
They share however some qualitative
features such as the fact that the
four derivative terms in the $N=2$ effective action derived
from the fivebrane theory are singular precisely
at the same points in the moduli space of vacua where 
the four derivative terms of the four dimensional
gauge theories are.
It seems therefore that although the eleven-dimensional
supergravity 
limit cannot provide a quantitative agreement
between the fivebrane theory
and the gauge theory, it can still be useful to extract qualitative
results. 

In order to use the fivebrane of $M$ theory to study
four-dimensional gauge theory in detail,
it seems that we need to get a better understanding of the fivebrane
theory when $R$ is small and the
Type IIA string theory is weakly coupled.
Such an understanding might be provided by the matrix
theory description
of M theory. There are of course
several difficulties in that direction.
We are interested in theories with four 
and eight supercharges and the corresponding
matrix descriptions have half of these 
amounts of supersymmetry. This will make the analysis of the quantum
corrections difficult.
Also, the realization of the brane
configurations
in matrix theory
is not straightforward
since, for instance,
we do not know how to describe a transversal fivebrane in
matrix theory.

Another curious point that demonstrates the need for a better
 understanding of the fivebrane theory is the decoupling argument.
Generally in supersymmetric gauge theories we can think about parameters
of the theory as vev's of background chiral superfields. This is 
useful in order to get a control
on  quantum corrections. 
 For instance, consider a Type IIA string theory compactified
on a Calabi-Yau 3-fold which leads to $N=2$
theory in four dimensions. Using the way the dilaton enters the
low-energy effective action and 
the decoupling between the $N=2$ vector multiplet and hypermultiplets,
one can show that
the structure of the vector multiplet moduli space does not receive stringy
corrections and can be easily computed.
On the other hand the hypermultiplet moduli space
does receive stringy corrections.
In our case we have one additional parameter $R$.
However, it is not clear in which multiplet
$R$ sits. In particular, we were able to use the fivebrane theory
to compute both the Coulomb and Higgs branch metric which implies
that our understanding of the decoupling argument in the Type IIA
compactification 
does not apply in an obvious way to the fivebrane theory.

\noindent
{\bf Acknowledgements}

We would like to thank 
G. Horowitz, J. Maldacena, H. Murayama and C. Vafa
for useful discussions. 
This research is supported in part by
NSF grant PHY-95-14797 and DOE grant DE-AC03-76SF00098.
JdB is a fellow of the Miller Institute for Basic Research in
Science.

\appendix{The Space-time Metric}

We identify parameters defining the branes with some physical
quantities.
Specifically, in the $N=2$ configuration, we identify the position
of the D4 branes in the 4,5 directions (parametrized by $v$)
as the VEV of the scalar component of the vector multiplet
$\Phi$,
while the position in the 7,8,9 directions (parametrized by
$\vec{w}=(w^7,w^8,w^9)$) as the meson VEVs $|Q|^2-|\tilQ|^2$, $\tilQ Q$.
The purpose of this note is to write down the metric of the
eleven-dimensional space-time in terms of these coordinates $v,
\vec{w}$.
In the flat back-ground, we claim that it is given by
\beq
\dd s^2=-(\dd x^0)^2+(\dd x^1)^2+(\dd x^2)^2+(\dd x^3)^2
+|\elst^2\dd v|^2+|\elel^3 \dd \vec{w}|^2
+\left|R\frac{\dd t}{t}\right|^2,
\label{11d}
\eeq
where $t$ is the complex coordinate for the 6,10 directions
given by $t=\exp(-x^6/R-ix^{10})$ ($x^{10}$ has period $2\pi$). 
In the above expression,
$\elst$ is the string length, $\elel$ is the eleven-dimensional
Planck length, and $R$ is the radius of the circle in the eleventh
direction which are related by $\elel^3=R\elst^2$.
In the presence of D6 branes at points in the 4,5,6
directions,
the metric is obtained by replacing
$|\elst^2\dd v|^2+|R\dd t/t|^2$ in (\ref{11d})
by the multi-Taub-NUT metric $\dd s_{\rm TN}^2$ which is
exhibited in some detail in Appendix B.

It is easy to see that the length of an interval in the $v$ direction
is given by $|\elst^2\Delta v|$.
Look at two D4 branes at $v=a_i$ and $v=a_j$ which are stretched
between NS 5-branes.
The string stretched between these D4 branes
creates the $(i,j)^{\rm th}$ component of the W-boson.
This W-boson has a mass $|a_i-a_j|$, while the stretched string
has a mass
given by string tension $\elst^{-2}$ times the length. 
Thus, the length of the string must be $\elst^2|a_i-a_j|$.

Let us next look at the D4 branes which are sliding
between D6 branes and are separated in the $\vec{w}$ direction.
It is not obvious with what to identify the states created by
strings stretched between them and,
therefore, the above argument does not apply
to measure the length of the separation in the $\vec{w}$ direction. 
However, we can compactify on a circle and T-dualize
to go to Type IIB theory, and apply the
S-duality. Then, the NS 5-branes become D5 branes,
D6 branes become NS 5-branes and
the D4 branes sliding between D6 branes become D3 branes
sliding between NS 5-branes, and we can identify the states
created by the string stretched between such D3 branes as the
W-bosons in the effective three dimensional gauge theory.
This three-dimensional theory is the mirror of
the theory obtained by compactification on
the circle of the original four-dimensional theory, and
the Higgs branch of the original theory is given by the
quantum Coulomb branch of this mirror theory.
Since we know how the moduli parameters are mapped under
the mirror symmetry, we can measure the length of the separation
in the $\vec{w}$ direction by expressing the same length
in terms of the parameters of the mirror gauge theory.

Let $R_{\rm A}$ be the radius of the compactification circle.
After T-duality and S-duality transformations,
we obtain Type IIB theory with the string tension
$\widetilde{\ell_{\it st}}^{-2}
=(\gst \elst^{3}/R_{\rm A})^{-2}$.
The squark $Q_3$ of the compactified theory is related to the
squark $Q$ of the four-dimensional theory by $Q_3=\sqrt{R_{\rm A}}Q$
and, under the mirror symmetry, the bilinear $|Q_3|^2$
is mapped to the
scalar component $\widetilde{\phi}$ of the vector multiplet
of the mirror.
Therefore the length of the separation
$\Delta \widetilde{\phi}=\Delta |Q_3|^2=R_{\rm A}\Delta |Q|^2
=R_{\rm A}\Delta w$ 
is given by
\beqa
\widetilde{\ell_{\it st}}^{2}|\Delta \widetilde{\Phi}|
&=&\gst \elst^3/R_{\rm A}\cdot R_{\rm A}|\dd\vec{w}|
\nonumber\\
&=&\elel^3|\dd\vec{w}|.
\eeqa

There is another test of the metric (\ref{11d})
by comparing the brane computation with the field theory result
concerning the domain wall of $N=1$ super-Yang-Mills
theory with gauge group $SU(n)$. Since the domain wall is
a BPS object, we expect that
the brane computation will correctly reproduce its
tension (For BPS mass formula, see \cite{Fayy,Piljin,Mikhailov}
for $N=2$ theory in four dimensions and
\cite{HH} for $N=2$ theory in two dimensions).
In \cite{hoo,W}, the brane configuration corresponding to the
$N=1$ super-YM theory is constructed by rotating the configuration for
the $N=2$ super-YM theory. The configuration for the theory with
the dynamical scale
$\Lambda$ is given by
\beq
vw=\Lambda^3,~~~
t=w^n
\label{sYM}
\eeq
The domain wall in the brane theory \cite{W,volo}
is given by a configuration of
the fivebrane which varies in the $x^3$ direction in such a way that
it approaches the configuration (\ref{sYM})
as $x^3\to -\infty$ while as $x^3\to+\infty$ it approaches
the configuration (\ref{sYM}) with $\Lambda^3$ being replaced
by $\Lambda^3\e^{2\pi i/n}$.
The tension is proportional to
the regularized volume of the non-trivial
part of the configuration, and is given by the integration of
the holomorphic three form times the fivebrane tension $\elel^{-6}$.
The holomorphic three form associated with the metric (\ref{11d})
is given by
\beq
\Omega=\elst^2\dd v\wedge \elel^3\dd w\wedge R{\dd t\over t}
=\elel^6\,\dd v\wedge\dd w\wedge {\dd t\over t}.
\eeq
The computation by \cite{W} leads to the tension formula
\beq
{1\over \elel^6}\left|\,\int\Omega\,\right|
=4\pi n |\Lambda^3(1-\e^{2\pi i/n})|,
\eeq
which agrees with the field theory result \cite{KSS}.
\footnote{This refers to the computation in section 5 of
\cite{KSS}
of the tension of
the domain wall separating two chirally asymmetric vacua
of the theory with massive quark multiplets where there
is a good description of the low energy theory.}

\appendix{The Taub-NUT Space}

In the presence of D6 branes at points in the 4,5,6
directions the space-time metric is obtained by replacing
$|\elst^2\dd v|^2+|R\dd t/t|^2$ in (\ref{11d})
by the metric of the multi-Taub-NUT space.
Here we collect some useful facts and formulae about this space
(see \cite{Hit,Nakatsu}).

If the D6 branes are at
$\vec{x}=(x^4,x^5,x^6)=\vec{x}_i$, $i=1,\ldots, N_f$,
the corresponding Taub-NUT metric
is given by
\beq
\dd s_{\rm TN}^2\,=\,
U\,\dd \vec{x}^2
+R^2\,U^{-1}\left(\dd x^{10}+\vec{\omega}\cdot\dd\vec{x}\right)^2
\label{TN}
\eeq
where
\beq
U\,=\,1+\sum_{i=1}^{N_f}\frac{R}{2|\vec{x}-\vec{x}_i|}\,,~~~
{\rm curl}\,\vec{\omega}={2\over R}{\rm grad}\, U\,.
\label{Uom}
\eeq
Since $x^{10}$ has period $2\pi$, it is a circle bundle over the
$\vec{x}$ space at large
$|\vec{x}|$ whose asymptotic radius is $R$. The equation
(\ref{Uom}) shows that $\vec{\omega}$
has magnetic charge $N_f$ and hence the circle bundle has first Chern
class $N_f$ on a two sphere at a fixed large $|\vec{x}|$.

This is a hyper-K\"ahler space.
Among the family of complex structures parametrized by a two-sphere,
we choose the one such that $x^4+ix^5$ is a complex coordinate.
 For reasons explained in Appendix A, we identify this with
$\elst^2 v$:
\beq
v=\elst^{-2}(x^4+ix^5)\,.
\eeq
Another complex coordinate has a more involved expression:
\beq
y=\e^{-(x^6/R+ix^{10})}\prod_{i=1}^{N_f}
\sqrt{|\vec{x}-\vec{x}_i|-(x^6-x^6_i)}\,\cdot\,{\rm const}\,.
\label{defY}
\eeq
In terms of these coordinates, the metric (\ref{TN})
is expressed as
\beq
\dd s_{\rm TN}^2=U\,|\elst^2\dd v|^2
+R^2\,U^{-1}|{\dd y/y}-\delta \dd v|^2
\label{TN2}
\eeq
where
\beq
\delta={1\over 2}\sum_{i=1}^{N_f}{1\over v+m_i}
\left(1+{x^6-x^6_i\over |\vec{x}-\vec{x}_i|}\right)\,,
\eeq
in which $-m_i$ are the locations of D6 branes in the $v$ directions,
that is, $m_i=-\elst^{-2}(x^4_i+ix^5_i)$.
If we introduce an another coordinate
\beq
x=\e^{x^6/R+ix^{10}}\prod_{i=1}^{N_f}
\sqrt{|\vec{x}-\vec{x}_i|+(x^6-x^6_i)}{v+m_i\over |v+m_i|}\,
\cdot\,{\rm const},
\label{defX}
\eeq
we find the equation
\beq
xy={\rm const}\cdot\prod(v+m_i).
\label{Asingu}
\eeq
If all $m_i$ are the same, it appears from this
that there is a singularity at $x=y=v+m=0$,
but it is actually smooth as long as $x^6_i$ are distinct..
This corresponds to the resolution of the $A_{N_f-1}$ singularity
as described e.g. in \cite{hov,hoo}.

Since it is a Ricci flat K\"ahler manifold with respect to the chosen
complex structure, there is a nowhere vanishing holomorphic two-form
$\Omega$ such that the square of the K\"ahler form $\omega$
is given by $\omega^2=\Omega\wedge\overline{\Omega}$.
It is
\beq
\Omega=\elel^3\,\dd v\wedge {\dd y\over y}.
\eeq
The coefficient $\elel^3$ is because of the relation
$\elst^2\cdot R=\elel^3$.
That this is nowhere vanishing is obvious in the case
where $m_i$ are distinct and (\ref{Asingu}) needs no resolution.
When $m_i$ are coincident (e.g. all zero), that is 
most easily seen by expressing this in each patch of the resolved
surface. In the description of \cite{hoo}, 
$\Omega=\dd x_i\wedge\dd y_i$ up to a constant
in the $i^{\rm th}$ patch coordinatized by $(x_i,y_i)$.

\end{document}